\documentclass[10pt,preprint]{aastex}

\shorttitle{Stellar wind structure of $\eta$ Carinae}
\shortauthors{Smith et al.}

\begin{document}

\title{LATITUDE-DEPENDENT EFFECTS IN THE STELLAR WIND OF ETA CARINAE\altaffilmark{1}}

\altaffiltext{1}{Based on observations made with the NASA/ESA {\it
Hubble Space Telescope}, obtained at the Space Telescope Science
Institute, which is operated by the Association of Universities for
Research in Astronomy, Inc., under NASA contract NAS5-26555.}

\author{Nathan Smith\altaffilmark{2} and Kris Davidson} 
\affil{Astronomy Department, University of Minnesota, 116 Church
St. SE, Minneapolis, MN 55455}

\altaffiltext{2}{Current address: Center for Astrophysics and Space
Astronomy, University of Colorado, 389 UCB, Boulder, CO 80309; {\tt
nathans@casa.colorado.edu}}

\author{Theodore R. Gull}
\affil{Laboratory for Astronomy and Space Science, NASA/Goddard Space
Flight Center, Greenbelt, MD 20771}

\author{Kazunori Ishibashi}
\affil{Center for Space Research, Massachusetts Institute of
Technology, 77 Massachusetts Ave., Cambridge, MA 02139}

\author{D.\ John Hillier}
\affil{Department of Physics and Astronomy, University of Pittsburgh,
3941 O'Hara Street, Pittsburgh, PA 15260}

\begin{abstract}

The Homunculus reflection nebula around $\eta$ Carinae provides a rare
opportunity to observe the spectrum of a star from more than one
direction.  In the case of $\eta$ Car, the nebula's geometry is known
well enough to infer how line profiles vary with latitude.  We present
STIS spectra of several positions in the Homunculus, showing directly
that $\eta$ Car has an aspherical stellar wind.  P Cygni absorption in
Balmer lines depends on latitude, with relatively high velocities and
strong absorption near the polar axis.  Stronger absorption at high
latitudes is surprising, and it suggests {\it higher mass flux toward
the poles}, perhaps resulting from radiative driving with equatorial
gravity darkening on a rotating star.  Reflected profiles of He~{\sc
i} lines are more puzzling, offering clues to the wind geometry and
ionization structure.  During $\eta$ Car's high-excitation state in
March 2000, the wind was fast and dense at the poles, with higher
ionization at low latitudes.

Older STIS data obtained since 1998 reveal that this global
stellar-wind geometry changes during $\eta$ Car's 5.5 year cycle, and
may suggest that this star's {\it spectroscopic events are shell
ejections}.  Whether or not a companion star triggers these outbursts
remains ambiguous.  The most dramatic changes in the wind occur at low
latitudes, while the dense polar wind remains relatively undisturbed
during an event. The apparent stability of the polar wind also
supports the inferred bipolar geometry.  The wind geometry and its
variability have critical implications for understanding the 5.5 year
cycle and long-term variability, but do not provide a clear
alternative to the binary hypothesis for generating $\eta$ Car's
X-rays.

\end{abstract}


\section{INTRODUCTION}

At visual wavelengths, the dusty bipolar ``Homunculus'' ejecta
structure around $\eta$ Car is primarily a hollow reflection nebula
(see Thackeray 1961; Visvanathan 1967; Meaburn 1987; Hillier \& Allen
1992; Hamann et al.\ 1994; etc.).  Therefore, by observing localized
parts of the Homunculus, {\it we can indirectly see the star's
spectrum from a range of directions in space} --- potentially allowing
us to reconstruct the shape of $\eta$ Car's wind, which is expected to
lack spherical symmetry.  Independent of the large-scale bipolar
morphology, hints of non-sphericity have been noted by, e.g., Viotti
et al.\ (1989), Hillier \& Allen (1992), Davidson et al.\ (1995),
Davidson \& Humphreys (1997), and Zethson et al.\ (1999).  Indeed,
Hillier \& Allen (1992) found a difference between direct and
reflected He~{\sc i} emission that originates in the stellar wind, and
early Space Telescope Imaging Spectrograph ({\it HST}/STIS)
observations showed that the reflected H$\alpha$ profile depends on
location within the Homunculus.

Departures from spherical symmetry are obviously critical for theories
of winds and instabilities in the most massive stars (see Lamers \&
Cassinelli 1999, and papers in Wolf, Stahl, \& Fullerton 1998).
Asphericity in $\eta$ Car's wind may also be critical for
understanding its 5.5 year cycle (Damineli 1996), its long term
variability, and excitation of its ejecta.  In addition to its classic
status as a very massive unstable star, $\eta$ Car is uniquely
favorable for this problem because the dusty Homunculus has a fairly
definite three-dimensional structure; each projected location on the
nebula provides reflected light from a viewing direction whose stellar
latitude is approximately known (Davidson et al.\ 2001; Smith 2002).

We have obtained long-slit STIS spectra optimized for studies of the
reflected latitude-dependent effects.  We find that some features do
in fact depend on viewing direction.  Wind velocities are highest
around the pole, as one might expect if stellar rotation reduces the
effective gravity and escape velocity at the equator.  Dominant
outflow velocities in the stellar wind are 400 to 600 km s$^{-1}$, but
some faster material ($\sim$1000 km s$^{-1}$) had been seen (Viotti et
al.\ 1989; Damineli et al.\ 1998), and our data now show that these
high velocities are polar.  We also find a much less anticipated
phenomenon: broad P Cyg absorption components of H$\alpha$ and
H$\beta$, usually weak in our direct view of $\eta$ Car, {\it are
quite deep at higher (not lower) latitudes}.  Traditionally, a slow
equatorial wind is supposed to be denser than the associated fast
polar wind and the mass-loss rate may increase with decreasing
$g_{eff}$ (Friend \& Abbott 1986); some examples are B[e] stars
(Zickgraf et al.\ 1986), wind-compressed disks or zones (Bjorkman \&
Cassinelli 1993; Ignace, Cassinelli, \& Bjorkman 1996; Owocki,
Cranmer, \& Blondin 1994), and bistable winds (Lamers \& Pauldrach
1991).  Our observations of $\eta$ Car, however, suggest that its wind
is densest near the poles despite higher velocities there.  Other
interpretations are conceivable, but require detailed models of line
transfer.

If $\eta$ Car's wind densities are highest near the poles as one might
guess from the deeper hydrogen P Cyg absorption, then this tends to
support recent comments by Owocki and others concerning the importance
of non-radial forces and gravity darkening in line-driven winds of
massive stars (Owocki, Cranmer, \& Gayley 1996; Owocki \& Gayley 1997;
Owocki, Gayley, \& Cranmer 1998; Maeder \& Meynet 2000; Meader \&
Desjacues 2001).  Specifically, these studies predict that if a
luminous star is rotating fast enough for gravity darkening to be
important (i.e., at $\ga$70\% of the ``critical'' rotation speed),
then stronger radiative flux at the hot poles will tend to drive a
higher mass flux.  Glatzel (1998) and Maeder (1999) have discussed
modifications to the ``critical'' rotation rate near the Eddington
luminosity.

We describe new STIS observations in \S 2 below, hydrogen line
profiles in \S 3, and helium profiles in \S 4.  In \S 5 we discuss
various likely implications of the observed wind geometry.  Then in \S
6 we examine older STIS spectra and discuss temporal variability of
the wind structure.  Finally, in \S 7 we mention consequences that a
variable, bipolar wind may have for interpretations of $\eta$ Car's
5.5-year spectroscopic cycle.

\section{STIS SPECTRA OF THE HOMUNCULUS}
 
These observations are part of an intensive {\it HST}/STIS project to
study $\eta$ Car and the Homunculus during their current 5.5-year
spectroscopic cycle (Davidson et al.\ 1999, 2000, 2001; Gull et al.\
1999; Gull \& Ishibashi 2001).  Early observations in March 1998 and
February 1999 used only a narrow 0.1$\arcsec$-wide STIS slit with
short exposures, inadequate for outer parts of the Homunculus
(although we did obtain a few deeper exposures at wavelengths near
H$\alpha$ that were offset from the star; see below).  Moreover,
difficulties in scheduling {\it HST} prevented the slit from being
aligned closely with the Homunculus bipolar axis during those earlier
observations.  Therefore, we planned a special set of STIS
observations with a wider 52$\arcsec\times$0$\farcs$2 aperture and
longer exposures, and with a slit orientation closer to the Homunculus
axis at position angle 138$\fdg$9$\leftrightarrow$318$\fdg$9, as shown
in Figure 1.  Here we discuss observations sampled with STIS gratings
G430M and G750M at standard tilts that gave central wavelengths of
4961, 6768, and 7283 \AA.  Achieved on 2000 March 13, these are the
same data used in a paper on the Homunculus shape (Davidson et al.\
2001).  In order to reduce scattered light we occulted the star with a
0.65$\arcsec$ bar across the slit, but other STIS observations gave a
direct view of the star a week later (see below).  Figure 2 shows
examples of long-slit STIS spectra for a few bright reflected emission
lines in $\eta$ Car's stellar wind (reflected continuum has been
subtracted).

To study positional variation of line profiles in the Homunculus, we
made several spectral extractions at positions along the slit listed
in Table 1.  Positive offsets are NW of the star, and negative offsets
are SE.\footnotemark\footnotetext{Table 1 lists information for
H$\alpha$ profiles, but the same extractions were made for other
emission lines discussed later.  Some offsets listed in figures for
other emission lines differ from those in Table 1, because adjacent
tracings were combined to increase signal-to-noise.}  The spectrum at
each offset position has a net redshift, due to reflection by
expanding dust.  We corrected for this effect by aligning the blue
side of the H$\alpha$ emission line profile at 10 times the continuum
flux (however, a different choice would not alter the results
substantially because the profiles match well). The net shift applied
at each position is given as $\Delta v$ in Table 1.  We used these
values of $\Delta v$ to correct the apparent velocities of other
emission lines discussed later.  Each resulting line profile samples
the spectrum of the central star seen from different positions in the
reflection nebula, and each corresponds to a different latitude as
depicted in Figure 3. This technique works because the polar lobes are
mostly hollow, as indicated by limb-brightening in thermal-infrared
images (e.g., Smith et al.\ 1998, 2002).

The case of $\eta$ Car is unique, because the geometry of the bright,
hollow reflection nebula is known sufficiently well that we can
correlate position in the nebula with stellar latitude (assuming that
the polar axis of the star and nebula are aligned -- in this case the
assumption is justified by the results, see \S 3).  Davidson et al.\
(2001) estimated the shape and orientation of the Homunculus from the
same dataset used here, and we use their ``compromise'' shape (shown
in Figure 3) to derive the latitude for each extraction position as
listed in Table 1 and depicted in Figure 3.  Since the Homunculus may
not be fully axisymmetric, our estimated latitudes have uncertainties
of roughly $\pm$5$\arcdeg$.  The axis for the southeast Homunculus
lobe is tilted at $i \; \approx \; 41\arcdeg \pm2\arcdeg$ relative to
our line of sight (see Davidson et al.\ 2001; Smith 2002).

The 2000 March 13 observations used a 0$\farcs$65 bar to occult the
central star, but other STIS spectra obtained on 2000 March 20 sampled
the star with a 0$\farcs$1-wide slit.  A 0$\farcs$15 segment was
extracted from the 2000 March 20 data for all emission lines discussed
here.  We also examined the spectrum of one position in the equatorial
ejecta located $\sim$5$\arcsec$ NE of the star, constructed from
several adjacent slit positions used to map the Homunculus at
wavelengths near H$\alpha$ on 2000 March 21.  The extracted spectrum
covers an effective area of 1$\arcsec$$\times$1$\arcsec$, represented
by the box labeled ``NE 5'' in Figure 1.  This spectrum is discussed
in \S 5.4 (and see Table 1).  Temporal variations should not introduce
significant errors since the datasets were taken only 1 week apart,
while differences in light travel times for reflected light from the
SE polar lobe are only a few weeks or less.  Thus, the observed
variation in line profiles {\it along the slit} cannot be due merely
to temporal variations in the spectrum.

In \S 6 we investigate temporal variability of reflected profiles as
seen in STIS spectra obtained with the G750M grating (central
wavelength = 6768 \AA) on 1998 March 19 and 1999 February 21.  These
are long-slit spectra similar to those obtained on 2000 March 13, but
were obtained using a 0$\farcs$1-wide slit located 0$\farcs$4 NW of
the star at position angle 152$\fdg$1$\leftrightarrow$332$\fdg$1, as
shown by the dotted line in Figure 1.  Transformation between position
along the slit and corresponding stellar latitude is less
straightforward than for the March 2000 data.  For the 1998 and 1999
data, we estimated latitudes for each offset position using contours
superimposed on images of the Homunculus (Davidson et al.\ 2001).
Uncertainty in latitude for these extractions is about
$\pm$5$\arcdeg$.  Spectra of the star for comparison with reflected
spectra were obtained simultaneously, and processed in the same manner
as stellar spectra discussed above.  In context with variability
during $\eta$ Car's 5.5 year spectroscopic cycle (Damineli 1996), STIS
observations in March 1998, February 1999, and March 2000 occur at
phases of 0.04, 0.21, and 0.4, respectively (see \S 5.1.2 and \S 7).

In \S 3 we discuss a small portion of the ultraviolet STIS FUV/MAMA
spectrum of $\eta$ Carinae, obtained in echelle mode with the E140M
grating on 2000 March 23, using a 0$\farcs$2 $\times$ 0$\farcs$2
aperture centered on the central star.  These data are used here to
help understand aspects of the optical spectra; a detailed analysis of
the UV spectra will be presented in a future publication.

\section{HYDROGEN LINE PROFILES}

\subsection{Latitudinal Dependence of P Cygni Absorption}

Figure 2 shows that reflected line profiles vary with position along
the slit.  The primary change occurs in strength and velocity range of
hydrogen P Cygni absorption, which is strongest 5$\arcsec$ to
6$\arcsec$ SE of the star.  P Cygni absorption is weak in the
directly-observed star and in reflected light from the NW lobe.
H$\beta$ and H$\alpha$ profiles agree well (Figures 4 and 5); their
scattering wings and emission peaks vary little across the SE lobe,
but absorption components depend on position.  Absorption traces a
more direct and narrow line of sight between reflecting dust in the
polar lobes and the star (this is particularly true for Balmer
absorption, which occurs at large radii compared to the continuum),
whereas emission comes from a range of
latitudes.\footnotemark\footnotetext{It is unlikely that the very
broad P Cygni absorption features arise in material within the polar
lobes or outside them, because intrinsic emission lines at those
locations have narrow line widths (see Davidson et al.\ 2001; Hillier
\& Allen 1992).}  Since each position along the slit translates to a
line of sight down onto a different latitude on the central star, P
Cygni absorption components give direct evidence that the density
and/or ionization structure of $\eta$ Car's current stellar wind is
non-spherical.  The possible cause of this aspherical structure is
addressed in \S 5.

To quantify the latitudinal dependence of velocity, we measured
``terminal'' speed $v_{\infty}$ and minimum-flux speed $v_{pcyg}$ for
each tracing in Figure 4.  Figure 6 shows how we defined these two
velocities observationally, and resulting measurements are listed in
Table 1.  With data from Table 1, Figure 7 indicates how observed
expansion velocities vary with latitude.  Changes in $v_{\infty}$ are
greater than the uncertainty, which could be up to 100 km s$^{-1}$ for
high latitudes.  For our line of sight, $v_{\infty}$ measured in
Balmer lines is within $\sim$50 km s$^{-1}$ of the terminal velocity
indicated by UV wind lines, like Si~{\sc ii} $\lambda$1527, also
observed in March 2000 using the STIS FUV/MAMA detector (Figure 8).
Several aspects of hydrogen emission-line profiles in Figures 4 and 5,
and trends in Figure 7 deserve comment:

1.  {\it Higher velocities near the pole.}  Blueshifted absorption is
seen at speeds up to $\sim$1100 km s$^{-1}$, which is much faster than
values usually quoted for $\eta$ Car's stellar wind.  The fastest
material between 5$\arcsec$ and 6$\arcsec$ SE of the star nearly
coincides with the polar axis of the Homunculus.  In Figure 7, the
measured quantities $v_{\infty}$ and $v_{pcyg}$ both follow a similar
trend of smoothly increasing speed with latitude, and there appears to
be a rather extreme increase in $v_{\infty}$ close to the polar axis.

2.  {\it Deeper absorption near the pole.}  Perhaps the most
significant and surprising result of this investigation is that the
deepest P Cygni absorption in Balmer lines is seen at the polar axis,
with absorption weakening {\it progressively} toward lower latitudes
(see Figure 5).  At latitudes below about 40$\arcdeg$, the normal
Balmer P Cygni absorption is absent in the
wind\footnotemark\footnotetext{This transitional latitude in the wind
  properties may have broader implications, since it corresponds to
  the approximate division between the caps and walls of the bipolar
  lobes (see Davidson et al.\ 2001; Smith 2002).}.  In dense winds
like $\eta$ Car's, the strength of P Cygni absorption in Balmer lines
is sensitive.  It depends on the population of the $n=2$ state and is
not necessarily a direct tracer of column density.  Weak Balmer
absorption at low to mid latitudes could be due to higher ionization
in that part of $\eta$ Car's wind, lower density, or a subtle
combination of both effects.

3.  {\it Symmetry about the equatorial plane and polar axis.}
Measured values of $v_{\infty}$ and $v_{pcyg}$ plotted together in
Figure 7 follow the same trend regardless of which polar lobe or which
side of the SE polar axis they correspond to.  Also, H$\alpha$
emission and absorption profiles on either side of the pole are
identical. This indicates that the wind outflow pattern is {\it
axisymmetric}.

\subsection{Enhanced Emission from the Star}

Electron scattering wings and line peaks seen at various positions in
the SE polar lobe all have the same profile shape (excluding
blueshifted absorption), but these reflected profiles all differ from
our direct view of the central star.  The difference is considerable
-- almost a factor of 2 in the peak intensity (Figure 5 has a
logarithmic scale).  Our direct view of the central star shows
stronger emission at velocities between about $-$250 and $+$500 km
s$^{-1}$ in both H$\beta$ and H$\alpha$, but broad scattering wings
are consistent with profiles seen in reflection.  Apparently, when we
look at the star we see extra line emission (or extra continuum
opacity) that is not seen by reflected lines of sight; the opposite
case would have been easier to explain.  A similar but lesser effect
appeared in the {\it HST}/FOS data on H$\beta$ (Davidson et al. 1995;
see also Hillier et al.\ 2001).  Later (in \S 6 and Figure 16) we show
that this characteristic of the star's emission persists at roughly
the same level throughout $\eta$ Car's spectroscopic variability
cycle.

Models of the central star's spectrum (as seen in March 1998) by
Hillier et al.\ (2001) showed a related problem: The model adequately
reproduced the electron scattering wings, but underestimated peak
emission of Balmer lines.  The model also overestimated the strength
of P Cygni absorption.  Thus, the atmosphere model used by Hillier et
al.\ approximates the reflected spectrum from the polar lobes in March
2000 more closely than it does the central star's spectrum, although
the observed outflow velocities at high latitudes exceed the 500 km
s$^{-1}$ terminal velocity adopted in their model.  This problem may
be related to wind geometry, since Hillier et al. assumed spherical
symmetry.

{\it Fe {\sc ii} Profiles.} Figures 9 and 10 show observed
emission-line profiles for Fe~{\sc ii} $\lambda$4925, extracted from
positions across the Homunculus and the central star in the same
manner as Balmer line profiles shown in Figures 4 and 5.  Fe~{\sc ii}
$\lambda$4925 is a bright permitted line in $\eta$ Car's wind, and
shows a broad profile similar to many other Fe~{\sc ii}
lines.\footnotemark\footnotetext{Fe~{\sc ii} $\lambda$4925 is blended
with He~{\sc i} $\lambda$4923, but the observed latitudinal behavior
of this line profile in comparison with other Fe~{\sc ii} and He~{\sc
i} lines suggests that it is dominated by the Fe~{\sc ii} transition.}
Fe~{\sc ii} and H$\alpha$ emission show a similar latitudinal
dependence.  Namely, the velocity and to some extent the strength of P
Cygni absorption increase with latitude.  Also, Fe~{\sc ii} emission
from the central star is stronger than at reflected positions.  Thus,
a solution to this problem for Balmer line emission should also
account for the Fe~{\sc ii} discrepancy.  More importantly, the
similar behavior of these two lines suggests that they arise in the
same outer zone of $\eta$ Car's wind, as expected because the
ionization of Fe~{\sc ii} is coupled to H by charge exchange in dense
stellar winds.

\section{HELIUM LINE PROFILES}

Figure 2 shows the long-slit spectrum across the major axis of the
Homunculus for He~{\sc i} $\lambda$7067 and $\lambda$6680.  Tracings
for both He~{\sc i} $\lambda$6680 and $\lambda$7067 at several
reflected positions across the nebula and for the central star are
shown in Figures 9 and 10.  Reflected line profiles in the SE lobe
seem to have three velocity components in Figure 2, caused by a
combination of a narrow central component from circumstellar gas, plus
a broad plateau or double-peaked line arising in the wind.

Contamination by reflected narrow He~{\sc i} $\lambda$7067 emission
from circumstellar ejecta like the Weigelt blobs (see Davidson et al.\
1995, 1997) needs to be examined before the wind lines can be
interpreted.  At 2$\arcsec$ to 4$\arcsec$ SE of the star, narrow
emission is conspicuous but the broad component and the continuum
emission fade.  This suggests that obscuring dust lies along this
particular line of sight to the star, projecting {\it a shadow} onto
the SE polar lobe.  This same position corresponds to a relatively
dark region in color {\it HST} images of the Homunculus (see Figure 1
and Morse et al.\ 1998).  Intervening material must be close to the
star (within $\sim$3500 AU) and compact ($r\lesssim$200 AU), because
it obscures the star along this line of sight but does not block the
Weigelt blobs.  The dust in question may be in condensations analogous
to the Weigelt blobs themselves.  Dense dust knots seen at
thermal-infrared wavelengths (Smith et al.\ 2002) are good candidates
for such blobs that are casting shadows.  Perhaps some other dark
spots in the Homunculus are also shadows cast by inner debris, rather
than holes or dense patches.

Changes in broad stellar-wind profiles are complicated, and may be
affected by a hypothetical companion star.  However, some clues to the
overall wind structure can be gathered.  Broad He~{\sc i}
$\lambda$7067 emission in the SE lobe (see Figure 9) has a nearly
symmetric profile. This broad profile, combined with the central
narrow component, creates the triple-peaked appearance of He~{\sc i}
$\lambda$7067 in the SE lobe in Figure 2.  The red side of He~{\sc i}
lines disappears in the reflected spectra from the NW polar lobe, in
the spectrum of the star, and in reflected spectra in the SE polar
lobe at positions near the star (see Figures 9 and 10).  The blue side
of the broad He~{\sc i} line profiles is much stronger in the NW lobe,
but roughly the same for the SE lobe and central star (Figure 10).  In
other words, low latitudes (below $\sim$50$\arcdeg$) seen from the NW
lobe have asymmetric profiles with extra blueshifted emission, whereas
high latitudes in the SE lobe show symmetric profiles and extra
redshifted emission compared to the central star.  It is more
difficult to ascertain the level of axisymmetry in He~{\sc i} emission
than for Balmer lines, because He~{\sc i} lines do not show such
drastic changes near the pole; He~{\sc i} emission does appear to be
symmetric about the equatorial plane.

The central star and reflected emission from the NW lobe (low
latitudes) also show P Cygni absorption at --400 km s$^{-1}$ in
He~{\sc i} lines, but the reflected spectra in the SE lobe (high
latitudes) do not.  Balmer lines show the opposite trend; namely,
stronger P Cygni absorption at high latitudes.  Important He~{\sc i}
lines at 10830 and 20581 \AA \ also show P Cygni absorption (Hamann et
al.\ 1994; Damineli et al.\ 1998; Smith 2001, 2002); it would be
interesting to see how these vary with latitude.  Continued monitoring
of direct and reflected He~{\sc i} line profiles during $\eta$ Car's
5.5 year cycle with STIS may provide a powerful diagnostic of the
stellar wind's complex ionization structure.

\section{WIND STRUCTURE DURING THE NORMAL HIGH-EXCITATION STATE}

Latitudinal variations of H and He~{\sc i} lines described above show
that the speed, density, and ionization in $\eta$ Car's wind are
non-spherical.  Do these indicate an inherently axisymmetric wind, or
is the structure influenced by a hypothetical hot companion star?  One
can imagine suitable parameters for a binary model, but in several
ways described below the observations more closely match theoretical
predictions for a dense line-driven wind from a rotating star (Owocki
et al.\ 1996, 1998; Owocki 2002).  Thus, we favor rotation of the primary
star as the dominant factor shaping $\eta$ Car's wind.

\subsection{A Non-spherical Stellar Wind?}

\subsubsection{Latitudinal velocity dependence}

\noindent Figure 7 shows maximum outflow velocities increasing rapidly
toward the poles.  If the terminal speed $v_{\infty}$ is proportional
to the escape speed as usually occurs in stellar-wind theory, and if
the surface rotation rate $\Omega$ does not perceptibly depend on
stellar latitude $\beta$, then one might expect a simple latitude
dependence:

\begin{equation} 
v_{\infty}({\beta}) \ \approx \ v_{\infty}({\rm pole}) \sqrt{ 1 - q^2 {\cos}^2 \beta }
\end{equation}

\noindent where $q \ \approx \ v_{\rm rot}/v_{\rm crit}$ and $v_{\rm
crit} \equiv \sqrt{GM/R}$ at the equator.  Higher polar velocities are
generally expected from any rotating star with lower effective gravity
at the equator.  An example of this trend is shown in Figure 7 for $q
= 0.9$, and obviously does not match the observed latitude dependence
exactly.  But this failure is not very surprising since the assumption
ignores possible complications; line-driven wind theory may be an
oversimplification for $\eta$ Car, opacity may depend on latitude, the
flow is not necessarily radial, etc.  However, the ratio of the polar
to equatorial velocities does seem marginally consistent with $q =
0.9$.

The rapid rise in $v_{\infty}$ toward the pole is somewhat
problematic, especially since no high-speed redshifted emission is
seen there.  Fast material may be concentrated in a narrow cone near
the polar axis, so it may not contribute significantly to the
H$\alpha$-emitting volume in the wind.  This must be addressed with
quantitative modeling.  A normal binary model offers no obvious
explanation for high polar wind velocities, especially since one
expects the orbital plane to be roughly aligned with the equator of
the star and Homunculus.  High velocities seen in reflected light from
the SE polar lobe give the first direct indication that the polar axis
of the Homunculus is aligned with the rotation axis of the central
star.  The alignment of these axes has important consequences for the
formation of the bipolar lobes and equatorial ejecta around $\eta$
Car.  Namely, it means that axial symmetry and ejection physics during
the Great Eruption may be directly linked to the star's rotation (see
\S 5.5).

High velocities seen in reflection also indicate fast-moving material
somewhere inside the Homunculus, but the absorption probably occurs
relatively close to the star and not in the middle of the polar lobes
or outside them (see \S 6.3).  Values for $v_{\infty}$ at high
latitudes are almost twice the expansion velocities for gas and dust
in the Homunculus (see Figure 7).  Therefore, we should expect
shock-excited emission inside the polar lobes as the stellar wind
overtakes slower material from the Great Eruption (Smith 2001, 2002;
Smith \& Davidson 2001).\footnotemark\footnotetext{In addition, there
appears to be some evidence for weak non-thermal radio emission from
the edges of the Homunculus (White 2002).}  The highest speeds we
observe in $\eta$ Car's wind and Homunculus are insufficient for
$\eta$ Car's hard X-rays, which presumably require shock velocities of
roughly 3000 km s$^{-1}$ (Corcoran et al.\ 1995; Ishibashi et al.\
1999; Pittard \& Corcoran 2002).

\subsubsection{Colliding winds and axisymmetry}

The weakness or absence of P Cygni absorption in Balmer lines at low
latitudes and the strange latitudinal behavior of He~{\sc i} lines
make a colliding-wind binary model worth considering.  A hot companion
star's wind might sculpt part of the dense primary wind (see, e.g.,
Pittard \& Corcoran 2002) and ionize its equatorial zones, but a
binary model does not give an obvious explanation for the fast polar
wind. If high speeds are induced by a companion star, we might expect
them primarily at low latitudes instead.

However, one argument against a companion star dominating the wind
structure is that the non-spherical wind appears to be {\it
axisymmetric}, as noted in \S 3.1.  In Figure 7, velocities from both
hemispheres and both sides of the polar axis follow the same smooth
latitudinal dependence, and P Cygni absorption in hydrogen lines has
the same depth on either side of the pole.  The evidence for
axisymmetry is not as clear in He~{\sc i} emission profiles, but they
are consistent with an axisymmetric wind.  If $\eta$ Carinae has a
companion star that causes colliding-wind X-ray emission, it must have
a very eccentric orbit (see contribution in Morse et al.\ 1999).  The
STIS observations described here were obtained in March 2000, when a
companion would be nearly at apastron.  The shock would be between the
two winds, while the far side of the primary star's wind should be
mostly unaffected -- a seriously asymmetric arrangement.

Therefore the basic stellar-wind geometry most likely results from the
star's rotation.  A hypothetical hot companion may help to explain the
He~{\sc i} emission and other effects, but this has not been demonstrated
and requires theoretical work.

\subsection{Density or Ionization?}

To understand the wind structure, we first need to recall how density
and ionization govern the behavior of P Cygni absorption in H$\alpha$
and H$\beta$.  Balmer absorption in dense stellar winds is sensitive
to changes in wind ionization structure, but that ionization structure
in turn depends sensitively on density.  Najarro, Hillier, \& Stahl
(1997) present a detailed analysis of the behavior of H and He~{\sc i}
lines in the spectrum of P Cygni.  Much of their analysis may be
applicable to $\eta$ Car's wind (see also Hillier et al.\ 2001).  They
find that in dense winds of hot stars, the appearance of Balmer lines
falls into one of two regimes, depending on H recombination.  Dense
winds that remain fully ionized show pure H emission profiles.  If
hydrogen recombines, however, the population of the $n$=1 state
increases dramatically, and so the Lyman series and Lyman continuum
become optically thick.  Because of high optical depth in Ly$\alpha$,
the $n$=2 state acts as a metastable level and gives rise to strong
Balmer absorption.  Essentially, Najarro et al.\ find that strong
Balmer absorption requires sufficient density to cause H recombination
in the outer wind.  This is evidently the case for $\eta$ Car's polar
wind; much weaker Balmer absorption toward low latitudes then suggests
that H remains mostly ionized there.  Najarro et al.'s calculations
show that even a small reduction in the mass-loss rate can prevent H
recombination; Hillier et al.\ (2001) comment that for $\eta$ Car,
dramatic changes in P Cygni absorption are seen if the mass-loss rate
in their model is lowered by factors of 2 to 4.  Thus, we can explain
increasing strength of Balmer absorption toward the pole if $\eta$
Car's wind has {\it a latitudinal density gradient, with higher
densities toward the pole}.  Evidently, the wind begins to reach a
critical level for recombination near latitudes of 40$\arcdeg$.  This
is analogous to (but opposite) the ``bistability'' model for B[e]
supergiants (Lamers \& Pauldrach 1991; Pauldrach \& Puls 1990).

Figure 11 shows a hypothetical picture of latitudinal structure in
$\eta$ Car's stellar wind during its normal high-excitation state.
The inner circle represents the stellar radius, which is somewhat
ambiguous because the wind is opaque (see Hillier et al.\ 2001,
Davidson et al.\ 1995). Figure 11 depicts two different zones in
$\eta$ Car's outer wind: 1) a high-density, low-ionization zone where
H recombines and produces Balmer absorption (shaded), and 2) a lower
density, high-ionization zone (hatched) producing the He~{\sc i}
emission, or at least relatively transparent so that more of the
underlying He~{\sc i} zone is seen, and where H remains ionized.  In
reality, of course, demarcation between these zones is not so sharp.
More theoretical work is needed to understand the observed line
profiles (especially He~{\sc i}), but this picture qualitatively
satisfies the main observational requirements of preceding sections:

{\it Polar View:} Seen from the pole, the spectrum shows deep,
high-velocity P Cygni absorption in Balmer and Fe~{\sc ii} lines
through dense regions of the recombined wind.  Looking through
relatively transparent extended He~{\sc i} emission zones on either
side of the star, an observer sees a symmetric He~{\sc i} profile,
with blue- and red-shifted velocities of similar intensity out to
about $\pm$400 km s$^{-1}$.  P Cyg absorption is weak or absent in
He~{\sc i} lines toward the poles, since there is no continuum source
behind these extended He~{\sc i}-emitting zones.

{\it Low Latitudes:} The low density and high-ionization at low
latitudes inhibit hydrogen P Cyg absorption in Balmer lines and favor
He~{\sc i} emission.  An observer will see stronger blueshifted
He~{\sc i} lines from deeper in the wind than other latitudes, since
redshifted He~{\sc i} emission is occulted by the star.

{\it Intermediate Latitudes (Our Line of Sight):} We see weaker
hydrogen P Cygni absorption than a polar observer because we see more
continuum from deeper in the hotter parts of the wind; for a polar
observer, all the continuum and electron scattering is behind the
absorbing region of the wind.  We also see mostly blueshifted He~{\sc
i} emission, but not as strong as for viewpoints at lower latitudes
because the He~{\sc i} is spread over a wider velocity range, and
because our line-of-sight is nearly along the transition between the
two zones.  The recombined polar region of the wind does not emit as
much He~{\sc i} and is responsible for the missing (or weaker)
blueshifted peak in the star's He~{\sc i} line profile as compared to
the NW lobe.

\subsection{Rotational Wind Compression}

The above results suggest that $\eta$ Car's wind has prolate mass
flux, with higher velocities, higher densities, and lower ionization
toward the pole.  This is surprising, because reduced effective
gravity at the equator of a rotating star might lead one to expect the
opposite; for example, rapidly rotating B[e] supergiants are thought
to have low-ionization equatorial density enhancements, as noted
earlier.  However, the prolate mass flux we observe in $\eta$ Car
during its normal high-excitation state is not without theoretical
precedent.

Considering only centrifugal forces in an expanding and rotating wind,
one might expect material to be deflected toward the equator (e.g.,
Bjorkmann \& Cassinelli 1993).  However, in a line driven wind other
factors are important; radiation-hydrodynamical simulations that
include non-radial forces with CAK (Castor, Abbott, \& Klein 1975)
formalism for line-driven winds show that equatorial density
enhancements can be inhibited (Owocki et al.\ 1996, 1998).
Furthermore, with gravity darkening, oblateness, and other effects,
Owocki et al.\ predict enhanced mass flux toward the polar directions
of a rotating star.  Analytic scaling laws from CAK theory (their
equation 46) can be modified to give an approximate relation (see
Owocki et al. 1996) for the dependence of local mass-loss rate on flux
$F$ and gravity $g_{eff}$, given by

\begin{equation}
\dot{M} \ \propto \ F^{\case{1}{\alpha}} \ g_{eff}^{1 - \case{1}{\alpha}}
\end{equation}

\noindent where $\alpha$ is the force multiplier power-law index that
is usually adopted in CAK theory. However, the prologue to von
Zeipel's theorem predicts that $F \; \propto \; g_{eff}$ for a
rotating star.\footnote{Some authors call the above relation ``von
Zeipel's theorem'', but in fact it is merely a preliminary to the
theorem itself, which dealt with a more complex problem.  See
Eddington (1926), Kippenhahn \& Weigert (1990), and von Zeipel
(1924).}  If this relation applies, if rotation is responsible for
$\eta$ Car's asymmetry, and if line-driven wind theory is a fair
approximation, then equation (2) implies that local mass flux
${\rho}v$ should be proportional to $g_{eff}$, which depends on
latitude.  Since $v$ tends to be more or less proportional to
$\sqrt{g_{eff}}$, we expect, very crudely, a wind density dependence
${\rho} \; \propto \; \sqrt{g_{eff}}$.  Thus, the polar wind density
can easily be higher than that near the equator.  Figure 7 suggests
that the ratio might be as large as 2, which, as noted in \S 5.2, may
allow deep P Cygni absorption features at the pole but not at low
latitudes.  Even if the above argument does not strictly apply, some
modifications to the theory for luminous stars tend to produce strong
polar mass flux (see, e.g., Maeder 1999; Maeder \& Desjaques 2001).
Here we have ignored obvious possibilities of latitudinal opacity
variations and of acceleration by the radiation continuum; a
quantitative investigation is merited but will be difficult.

The essential requirement of our proposed interpretation of $\eta$
Car's wind asymmetry -- that higher densities toward the pole lead to
recombination at large distances in the wind (as discussed by Najarro
et al.\ 1997) -- seems theoretically plausible if $\eta$ Car is
rotating rapidly enough to cause equatorial gravity darkening (see
Owocki et al.\ 1996, 1998; Owocki \& Gayley 1997).\footnote{One should
bear in mind, however, that the observed H$\alpha$ and H$\beta$
absorption features represent the NLTE population of the $n = 2$
level, which depends on various subtle factors.  Some of these, such
as three-dimensional velocity gradients, are not treated adequately in
any published wind models.}  Gravity darkening on the star would not
necessarily contradict the higher ionization observed at low/mid
latitudes in the wind, because that latitude-dependent ionization is
regulated by the local density instead of an anisotropic UV source (\S
5.2).  For instance, latitudes at $\sim$25$\arcdeg$ from the equator
should still ``see'' the hot pole of the star.  Material much closer
to the equator will have a more difficult time getting UV radiation
from the hot poles, leading one to expect lower ionization there; this
may indeed be the case, as discussed in the next section.

While higher polar velocities are a generic property of winds from
rotating stars, higher polar mass flux is a distinctive property of
line-driven winds with gravity darkening.  Thus, our observations
appear to support the hypothesis that $\eta$ Car's present-day stellar
wind is in fact radiatively driven.  This may not hold true during the
Great Eruption.

\subsection{An Equatorial Disk?}

Several authors have discussed evidence for strange kinematics in
$\eta$ Car's equatorial ejecta (Zethson et al.\ 1999; Smith \& Gehrz
1998; Morse et al.\ 2001; Davidson et al.\ 2001; Davidson \& Humphreys
1997).  Not all equatorial structures are coeval with the polar lobes
ejected during the Great Eruption. The Homunculus' ragged {\it
equatorial\/} structure distinguishes it from other bipolar nebulae.
Equatorial outflow in the present-day wind may be important for
understanding this larger-scale morphology.

Figure 4 includes the equatorial H$\alpha$ profile observed
$\sim$5$\arcsec$ NE of the central star\footnotemark\footnotetext{
This location unambiguously represents equatorial ejecta, since it is
not projected against the NW Homunculus lobe in images (see Figure 1).
Some other apparently equatorial features superimposed on the lobe may
be illusory.  For instance, the ``fan'' a few arcsec northwest of the
star is really a hole in the equatorial extinction, allowing that part
of the NW lobe to appear brighter (Smith et al.\ 1999, 2002).  Note
that the STIS spectrum at NE5 does not include the entire 1{\arcsec}
box (see Table 1).} (see Figure 1), and Table 1 lists values of
$v_{\infty}$ and $v_{pcyg}$ for this feature.  These outflow
velocities are higher than the expected trend at the equator, and deep
P Cyg absorption there contradicts the observation that blueshifted
absorption weakens with decreasing latitude.  Thus, the reflected
spectrum from NE 5 suggests that these trends reverse at the equator
--- perhaps indicating the presence of a dense equatorial disk. NE 5
might be peculiar since it is projected along a line from the star to
the NN ``jet'' (see Morse et al.\ 1998; Meaburn et al.\ 1993).
However, other STIS data obtained in November 1998 (not shown), with
the slit crossing other extended parts of the equatorial skirt, also
show obvious P Cygni absorption at the equator. Binary models that
invoke a strong wind from a companion star (e.g., Pittard
\& Corcoran 2002) would seem to impede the survival of such a thin
equatorial disk.

Perhaps some material escaping from mid-latitudes on the star is
deflected toward the equator -- in which case the same process may
cause both depletion of the outer wind density at mid-latitudes and
increased density near the equatorial plane.  Various theoretical
ideas including centrifugal wind compression and magnetic fields may
produce an outflowing equatorial disk in this way (Bjorkman \&
Cassinelli 1993; Washimi \& Shibata 1993; Balick \& Matt 2001;
ud-Doula \& Owocki 2002; Lee, Saio, \& Osaki 1991).  Alternatively,
low ionization in the ``disk'' may be related to the star's gravity
darkening (\S 5.3).

\subsection{Shape and Structure of the Homunculus}

The bipolar geometry depicted in Figure 11 naturally leads one to
wonder if the present-day stellar wind's bipolarity on small scales
(within $\sim$100 {\sc AU}) is related to the shape and structure of
the Homunculus Nebula on much larger scales.  Conditions in $\eta$
Car's wind during the Great Eruption were very different than today,
but let's assume for the moment that the geometry of the mass loss was
similar then.

Suppose that latitudinal dependence on density in $\eta$ Car's wind
160 years ago resembled Figure 11, but that during the eruption the
mass-loss rate increased by a factor of 100 for 20 years --- turned on
and off like a faucet.  In that case, most mass would have been
ejected toward the poles during that time, forming two thick polar
caps. Their radial thickness would be roughly $v \; \times \; 20 \;
{\rm yr} \; \approx \; 2500 \; {\rm AU} \; \approx \; 1\arcsec$. This
simple estimate matches observations; the Homunculus polar caps are
indeed roughly 1$\arcsec$ thick at radii around 8$\arcsec$ from the
star (Smith et al.\ 1998, 2002; Davidson et al.\ 2001).

This outburst scenario differs from previous hydrodynamic models for
the formation of the Homunculus that invoke wind interactions to
explain the observed bipolarity (e.g., Frank et al.\ 1995, 1998;
Dwarkadas \& Balick 1998; Langer et al.\ 1999).  We should not expect
post-eruption hydrodynamic effects to dominate the structure of the
Homunculus, since material ejected during the eruption was a factor of
100 more dense but had comparable velocities to the pre- and
post-eruption winds.  Ram pressure in the post-eruption wind is
insufficient to dominate the shape of the nebula or cause significant
acceleration. Proper motions of the Homunculus ejecta (Morse et al.\
2001; Smith \& Gehrz 1998; Currie et al.\ 1996) indicate linear
expansion.  A bipolar eruption of a rotating star seems the best
explanation for the basic Homunculus morphology (see Owocki \& Gayley
1997; Maeder \& Desjacques 2001; Dwarkadas \& Owocki 2002).

Small-scale structure in the Homunculus lobes may also be related to
details in the eruptive wind.  Davidson et al.\ (2001) noted that the
smallest observed feature size, about 0.1$\arcsec$ (see Figure 1 and
Morse et al.\ 1998), corresponds to the relevant speed of sound
multiplied by the time elapsed since the Great Eruption.  The observed
characteristic scale size is larger, but does not contradict this
basic idea: that the clumpy (``granular'') small-scale structure in
each bipolar lobe represents features in the eruptive wind that were
frozen into the freely expanding ejecta.  Given the quantitative
details mentioned above, the corrugated structures probably did {\it
not\/} result from an instability in the flow long after the eruption
(Kelvin-Helmholtz, Rayleigh-Taylor-like, or their kin).  The
structure, which pictorially resembles solar granulation, may have
resulted from wind instabilities or from effects near the stellar
surface (the sonic point).  According to Shaviv (2000), porous
structure should develop in a super-Eddington wind like that of $\eta$
Car during 1838--1858.  This porosity may be imprinted in clumps or
holes seen in the polar lobes, and may prove more useful than the
large-scale lobe shapes for understanding the Great Eruption itself.


\subsection{Related Problems}

The proposed wind structure with high ionization at low latitudes and
dense optically-thick low-ionization zones near the poles helps
explain several peculiarities:

1.  High-ionization [Ne~{\sc iii}] and [Fe~{\sc iii}] emission regions
exist a few hundred AU from the star even though the dense wind is
expected to absorb the star's ionizing photons (e.g., see model in
Hillier et al.\ 2001).  This is one motivation for the binary
hypothesis, wherein a second star can provide ionizing radiation.
However, the non-spherical wind that we propose may allow UV photons
to escape from the primary star at low latitudes.  This occurs despite
gravity darkening at the stellar surface; the wind is opaque near the
poles but semi-transparent at low latitudes.  The ``Weigelt blobs,''
bright emission-line sources northwest of the star, are indeed thought
to be located near the equatorial plane (Davidson et al.\ 1995, 1997).

2.  The brightness of the Weigelt blobs is also a problem, since they
seem to reflect more of the star's light than they intercept, for
their presumed size (see Davidson \& Humpherys 1986).  Dust is
continually forming in $\eta$ Car's wind, evidenced by the unresolved
hot dust in the core of the Homunculus seen in infrared images (e.g.,
Smith et al.\ 1998, 2002).  If dust forms primarily in dense polar
regions of the wind, it might help obscure the star along our line of
sight more than toward the Weigelt blobs.

3.  The simultaneous existence of separate high- and low-excitation
regions in the wind helps explain the range of spectral types from B2
to B8 Ia seen in the far-UV spectrum of $\eta$ Car by Ebbets et al.\
(1997).

4.  The spherical wind model used by Hillier et al.\ (2001)
overestimates the depth of the P Cygni absorption components in Balmer
and \ion{Fe}{2} lines.  Figures 4 and 10 show that both species show
deeper absorption in reflected spectra from the SE lobe than in our
direct line of sight to the star.  This supports the hypothesis (point
number 2 above) that dense polar zones of $\eta$ Car's wind dominate
the optical spectrum, and the spherical model used by Hillier et al.\
more closely approximates polar wind conditions.

5.  In addition to the Weigelt blobs, extended {\it equatorial} ejecta
around $\eta$ Car show peculiar excitation that might be less
mysterious if $\eta$ Car has a stronger UV flux at low latitudes.
Equatorial regions would also see enhanced Ly$\alpha$ emission
escaping from the polar wind, relevant to some proposed excitation
mechanisms (Johansson \& Hamann 1993; Hamann et al.\ 1999).  Notable
extended equatorial features are the ``purple haze'' in {\it HST}
images (Morse et al.\ 1998), gas in the inner torus seen in infrared
and optical images (Smith et al.\ 1998, 2002), the bizarre \ion{Fe}{2}
lines near $\sim$2507 \AA \ (see Davidson et al.\ 1995, 1997;
Johansson et al.\ 1998; Hamann et al.\ 1999; Johansson \& Letokhov
2001), and other \ion{Fe}{2} and metal lines in the equatorial skirt
with strange velocities and excitation (Zethson et al.\ 1999).
Extended equatorial gas in the circumstellar torus also shows peculiar
{\it variability} (Duncan et al.\ 1995; Smith et al.\ 2000; Smith \&
Gehrz 2000), which may be related to spectroscopic changes discussed
below.

\section{TEMPORAL VARIABILITY}

So far we have discussed the wind during its normal high-excitation
state as observed in March 2000, but $\eta$ Car also has occasional
``spectroscopic events'' in which the spectrum temporarily changes for
reasons that are not yet understood.  Originally these events were
interpreted as single-star shell ejections (Zanella, Wolf, \& Stahl
1984; Bidelman et al.\ 1993), but a companion star seems likely, given
the 5.5-year periodicity associated with the events (Damineli 1996;
Whitelock et al.\ 1994; Corcoran et al.\ 1995; Duncan et al.\ 1995;
Damineli et al.\ 1997, 1998; Davidson 1999).

Limited HST/STIS data were obtained during the most recent
spectroscopic event in 1997--98 and at a few later times.  Davidson et
al.\ (1999) described some results for the star viewed directly, and
remarked that the spectroscopic changes resembled a shell ejection
(see also Gull et al.\ 1999; Davidson 1999, 2001).  We can apply the
same latitude-dependent reflection method described above to the March
1998 and February 1999 STIS data, where H$\alpha$ and He~{\sc i}
$\lambda$6680 were sampled with long exposure times on a particular
slit orientation and placement shown in Figure 1.  Unfortunately this
slit position was not the same as for the March 2000 observations
discussed above, but it crossed the Homunculus in a similar way.  Here
we describe significant changes in the reflected spectra; a report on
the directly-viewed star is in preparation (Davidson et al.).

\subsection{He~{\sc i} Variability}

Figure 12 shows grayscale representations of long-slit spectra for
He~{\sc i} $\lambda$6680 in March 1998 and February 1999, and Figure
13 demonstrates variability of He~{\sc i} line profiles with tracings
of the central star, as well as reflected spectra from the NW and SE
lobes.  Equivalent widths measured for these tracings are plotted in
Figure 14.  At positions in the NW polar lobe, He~{\sc i}
$\lambda$6680 is absent in March 1998 and extremely weak in February
1999. Positions in the NW lobe during March 1998 reflect the spectrum
of the star 1 to 3 months earlier (soon after the event), and show
that this feature nearly vanished, at least at these low latitudes.
STIS spectra of the central star during the event in late December
1997 show that the He~{\sc i} $\lambda$6680 flux remained detectable
along our direct line of sight (Figure 13).  In the SE polar lobe the
equivalent width in Figure 14 is nearly constant. Also, in March 1998
the He~{\sc i} emission line in the SE lobe is stronger than it is in
the NW polar lobe almost a year later in February 1999.  This could
not be the case if changes in the He~{\sc i} line were spherically
symmetric (the time delay for positions in the SE lobe is only a few
weeks).

In \S 4 we noted that the SE lobe in March 2000 showed a broad,
roughly symmetric stellar-wind He~{\sc i} profile, plus reflected
narrow emission from slow-moving inner ejecta such as the Weigelt
blobs.  The narrow line disappeared during 1998 and early 1999;
compare Figure 12 to Figure 2.  Meanwhile, as shown in Figure 13, the
polar view of the stellar wind component (``SE lobe'') changed in a
way quite unlike that observed directly on the star.  The latitude
dependence of He~{\sc i} variability probably indicates that {\it the
wind's asphericity was changing\/} during and after the 1997--98
event.  Figure 14 shows that broad He~{\sc i} emission faded mainly at
low latitudes, while the polar wind seemed relatively undisturbed.

\subsection{H$\alpha$ Variability}

H$\alpha$ emission seen directly in the spectrum of the central star
has remained roughly constant in brightness relative to the continuum
and in width, but its P Cyg absorption changed during 1998--99 (see
Davidson et al.\ 1999; Davidson 2001).  Changes in the latitudinal
dependence of reflected P Cygni absorption are striking (Figure 12);
tracings of the H$\alpha$ line are shown in Figures 15 and 16, and
corresponding velocities are listed in Table 2. In our direct view of
the star, deep P Cyg absorption around $-520$ km s$^{-1}$ occurred in
early 1998 during the event but virtually disappeared before February
1999.  Meanwhile, however, strong P Cyg absorption persisted at high
latitudes, and its extreme velocity grew from $-650$ km s$^{-1}$ in
1998 to $-1000$ km s$^{-1}$ in 2000 (SE lobe, Figures 15 \& 16).  Deep
P Cyg absorption seen at low latitudes in March 1998 disappeared two
years later.

Thus, viewing the spectra seen in early 1998 rather than 2000, we
realize that {\it the strong latitude dependence described in \S 5
above did not apply during the 1997--98 spectroscopic event.\/} At
that time the depth of the P Cyg absorption was less
latitude-dependent, while the extreme observed speeds that we have
denoted $| v_{\infty} |$ temporarily decreased at high latitudes but
increased at low latitudes (Figure 17).  The wind was evidently
disrupted through a wide latitude range, most likely by a ``shell
ejection'' that increased the density especially at low latitudes.
This may help to explain why Hillier et al.\ (2001) could reproduce
the early-1998 spectrum fairly well with a spherical wind model, but
found later spectra more problematic.

\subsection{Implications for the Wind Geometry}   

In two ways the temporal variability appears very consistent with the
hypothetical bipolar wind geometry sketched in Figure 11 (\S 5).
First, a moderate shell ejection can easily decrease the ionization in
the low-latitude wind, but should have less effect on the polar wind
which is normally denser and less ionized.  Qualitatively, at least,
the observed behavior matched this expectation.  Second, the 1998--99
temporary disappearance of the highest observable velocities
($\sim$1000 km s$^{-1}$, Figure 5) suggests that the fast material
occurs in the polar wind {\it per se\/} within 0.5$\arcsec$ of the
star, and not at larger radii in the Homunculus lobes.  Our reason for
saying this is that the recombination time $(\alpha_{B} n_e)^{-1}$ is
too long for rapid response, i.e., longer than a year, at radii more
than a few hundred AU from the star.
     
Finally, we note that if the March 2000 data -- midway through the
5.5-year spectroscopic cycle -- represent the ``normal'' state, then
Figures 16 and 17 indicate that the post-event recovery extended over
more than a year.  This is not what one would expect in a simple
binary model, since the hypothetical companion star should have moved
a considerable distance, $\sim$10 AU, from the primary in less than 6
months after periastron passage (see, e.g., Davidson 1999).

\section{IMPLICATIONS FOR THE 5.5 YR SPECTROSCOPIC CYCLE AND A MORE LIMITED ROLE FOR THE HYPOTHETICAL COMPANION STAR}

The results discussed above suggest that the 1997--98 spectroscopic
event was essentially a disturbance of the primary wind, and not, for
instance, an eclipse or occultation.  Most likely the term ``shell
ejection'' is valid to some extent, as advocated by Zanella et al.\
(1984), Davidson (1999), and Davidson et al.\ (1999). None of the
observations directly requires a companion star; a single-star model
remains conceivable, but a binary model is not precluded either.
Here, largely independent of whether the system is a binary, we
describe a qualitative scenario for a spectroscopic event with the
observed latitude dependences, and we list a variety of ways in which
it appears consistent with observations.  We assume that stronger
hydrogen P Cygni absorption implies higher local mass density as
discussed in \S 5; other interpretations may exist but would be more
complicated.  Detailed calculations are obviously needed but will be
quite difficult, since a proper model must be two- or
three-dimensional, with continuum- and line-driven gas dynamics,
rotation, realistic NLTE level populations and radiative transfer,
etc.

\subsection{A Shell Ejection During the Spectroscopic Event}  

Figure 18 shows the likely arrangement of zones in $\eta$ Car's wind
at three sample times: before, during, and soon after a spectroscopic
event.  This is merely a schematic diagram, and boundaries between
zones are not necessarily sharp.  Figure 18$a$ usually applies:
ionizing UV can escape through the lower-density low-latitude regions
as conjectured in \S 5.  Then, during an event (Figure 18$b$), the
low-latitude wind becomes denser, mostly non-ionized, and opaque to
ionizing photons.  A density-increase factor of the order of 2 may
suffice and the extra mass ejected might be as little as $2 {\times}
10^{-5}$ $M_{\odot}$.  Meanwhile, a relatively weaker high-latitude
increase has less effect on the emergent spectrum; lessened values of
$v_{\infty}$ and $v_{pcyg}$ shown in Figures 16 and 17 may indicate
that hydrogen recombines closer to the star, or that the polar wind
slows as its mass increases.  A few months later (Figure 18$c$), the
ejected ``shell'' gradually becomes transparent.  Altogether this
scheme, where changing density at low latitudes acts as a sort of {\it
valve} regulating the escape of UV photons, can account for nearly all
observed characteristics of each spectroscopic event:

1.  Narrow [Ne~{\sc iii}], [Fe~{\sc iii}], He~{\sc i}, and related
emission lines temporarily fade -- the phenomenon that led Zanella et
al.\ (1984) to propose a UV-quashing ``shell ejection.''  We now know
that these features originate in slow-moving {\it low-latitude\/}
ejecta close to the star (Davidson et al.\ 1995, 1997; Smith et al.\ 
2000).

2.  Broad He~I emission lines weaken during an event (Damineli 1996;
Damineli et al.\ 1997, 1998).  These probably originate in
low-latitude wind zones (\S 6.1).

3.  Broad (i.e., wind) components of some low-excitation lines such as
Fe~{\sc ii} become stronger (Damineli et al.\ 1998).  This is
qualitatively unsurprising if the volume and mass of low-excitation
material increase as the wind becomes quasi-spherical (Figure 18$b$).

4.  In our direct view of the star, many violet and UV absorption
features deepen during an event, reminiscent of a shell ejection
(Davidson et al.\ 1999).

5.  In the time period between spectroscopic events, colliding-wind
X-ray binary models for $\eta$ Car (Ishibashi et al.\ 1999; Corcoran
et al.\ 2001; Pittard \& Corcoran 2002) seem to require a primary
mass-loss rate of only a few times 10$^{-4}$ $M_{\odot}$ yr$^{-1}$, a
factor of three to ten less than the familiar value based on other
considerations (Hillier et al.\ 2001; Davidson \& Humphreys 1997).
However, if the primary wind is not spherical in its ``normal'' state,
then the X-ray parameters refer mainly to local densities around the
shock interface, presumably within the low-latitude, lower density
part of the primary wind sketched in Figure 18$a$.  Corcoran et al.\
(2001) also invoke a sudden density increase to explain the collapse
of the X-ray flux during an event, which seems consistent with the
shell-ejection scenario depicted in Figure 18.

6.  The radio-wavelength behavior of $\eta$ Car includes varying
free-free emission near the equatorial plane (Duncan et al.\ 1995).
The source is compact during an event, and later expands irregularly
as ionizing photons penetrate to larger distances from the star.
Near-infrared images, also representing free-free continuum emission
mostly at low latitudes, show changes morphologically similar to the
radio emission (Smith \& Gehrz 2000).

7.  The 7 mm flux, produced by free-free emission in outer parts of
the wind, fades during a spectroscopic event (Abraham \& Damineli
1999).  This can be explained if some normally ionized (low-latitude)
portions of the wind recombine during the event.
(Millimeter-wavelength continuum mostly originates within 100 AU of
the star, while centimeter wavelengths mentioned above in point 6
represent ejecta farther out.)

8.  The variable geometry and emitting volume of the wind may cause
serious line profile variations like those that are observed.  As
noted by McGregor et al.\ (1999) and Davidson et al.\ (2000), this
makes it difficult or impossible to correlate radial velocities with
possible orbital motion.

9.  There is almost no obvious large-scale photoionization in the
high-latitude zones around the star; the [Ne~{\sc iii}] and similar
features mentioned in point 1 above appear to be fairly close to the
equatorial plane based on their Doppler velocities, and so is the
centimeter-wavelength emission mentioned in point 6.  Since there is
no shortage of gas at various radii within the Homunculus lobes (see
Ishibashi et al.\ 2001), we conclude that ionizing UV photons are
scarce at high latitudes.  This may be an objection to the most likely
type of binary model, wherein a hot companion star produces ionizing
radiation.  Since the primary wind does not seem dense enough to block
ionizing photons at the hypothetical apoastron distance ($\sim$ 30
AU), we should expect to see bright photoionized material in
near-polar locations around the star, contrary to observations.  The
single-star geometry sketched in Figure 18$a$, on the other hand, can
emit ionizing radiation toward low latitudes only, consistent with the
observed large-scale ionization.

\subsection{Rotation and a Companion Star's Influence}

In most binary schemes for $\eta$ Car's 5.5-year cycle (e.g., Damineli
et al.\ 1997, 2000; and various authors in Morse et al.\ 1999), a hot
companion star in an eccentric orbit is the main perpetrator.  It
provides a fast wind needed for the X-rays, and -- far greater in the
energy budget -- it supplies ionizing UV radiation to excite emission
lines in surrounding ejecta.  Emergent UV and X-rays are modulated by
the secondary star's varying radial location in the dense primary
wind.  In the latitude-dependent shell ejection scenario that we
propose, however, much (perhaps nearly all) of the UV radiation comes
from the primary rather than the secondary star.  A companion star may
still be needed for the hard X-rays from $\eta$ Car (Corcoran et al.\
1998, 2001; Ishibashi et al.\ 1999; Ishibashi 2001; Pittard \&
Corcoran 2002; etc.), but at wavelengths other than X-rays -- i.e.,
for 99.99\% of the energy flow -- the main role of a hypothetical
secondary star is to regulate the periodicity.  It may supply the most
extreme UV photons, but this is unproven.

Most likely a spectroscopic event occurs near periastron passage when
the secondary star tidally disturbs the surface layers of an already
unstable primary.  This is feasible, at a separation more than 3 times
the primary's radius, only because radiation pressure and (probably)
rotation magnify the tidal effect.  In \S 5.3 we noted how $\eta$
Car's wind geometry qualitatively matches theoretical predictions by
Owocki et al.\ (1996) for a rotating star with gravity darkening.  In
that formulation, gravity darkening followed $F \ \propto \ g_{eff}$
(von Zeipel 1924), and simple scaling arguments would lead to
$v_{\infty} \propto T^2$.  This is probably an oversimplification for
$\eta$ Car's optically-thick wind, but the agreement between the
observed wind structure and the associated predictions from theory
suggest that it is not too far-fetched. Therefore, the pole-to-equator
velocity ratio of $\sim$2 (Figure 7) during the normal high-excitation
state implies, very crudely, that $T_{pole}/T_{equator} \approx
\sqrt{2}$.  Thus, if $\eta$ Car's polar temperature is close to the
often-adopted value of 30,000 K (e.g., Davidson \& Humphreys 1997;
Hillier et al.\ 2001), then the corresponding temperature at low
latitudes could be dangerously close to 21,000 K, where an important
opacity increase toward lower temperatures occurs (see Pauldrach \&
Puls 1990; Lamers \& Pauldrach 1991).  Intuition then guides us to
ask: What happens if the rotation rate is slightly increased?  Such a
change should increase the amount of gravity darkening, decrease $T$,
and raise opacity at the equator --- i.e. rotation may trigger an
equatorial shell ejection.

The rotation in question may be innate or it may have been forced by a
companion star, but either way the bipolar Homunculus strongly
suggests that rotation is indeed significant in $\eta$ Car.  Thus we
conjecture that {\it systematic variations of the surface rotation
rate play a role in $\eta$ Car's 5.5-year cycle.\/} The reasoning,
mentioned earlier by Davidson (1999), goes like this:

1. Imagine, for simplicity, that $\eta$ Car initially has solid-body
rotation, with a constant angular rate $\Omega$.  (This assumption
merely makes the next two steps easier to visualize.)

2. Then any surface eruption carries away a larger-than-average
angular momentum per unit mass.

3. Afterward, the star very quickly readjusts to a new state of
quasi-hydrostatic equilibrium.  The dense inner region containing most
of the mass is scarcely affected, but the outermost surviving layers
must expand appreciably outward and therefore now have slower rotation
rates.

4. At this point the star's dense center rotates faster than the less
massive outer layers.  Therefore angular momentum will diffuse
outward, gradually accelerating the surface rotation.  According to
simple order-of-magnitude turbulent diffusion estimates, a spin-up
timescale of a few years seems plausible following the ejection of
less than 10$^{-3}$ $M_{\odot}$.

5. Thus we conjecture that the surface rotation rate gradually
increases during the 5.5-year period, with possible effects on the
low-latitude temperature (gravity darkening), on the wind, etc.
Rotation is always far slower than the naive ``break-up'' limit, but
toward the end of the cycle it may be fast enough to have a
destabilizing effect as noted earlier.  Internal thermal readjustment
can meanwhile play a related role.

6.  Most likely this development affects a binary model, and tidal
friction near periastron may play a long-term role.  However, one can
also imagine a {\it single-star\/} cycle based on the above reasoning.
In such a model, 5.5 years is the time required for spin-up to permit
another mass ejection event.  There is no obvious physical reason why
such a periodicity cannot be as regular as the observed cycle.  Other
single-star alternatives for short-term periodicity exist as well
(e.g., Stothers 2000; Guzik et al. 1999; Davidson 1999).  These could
act independently, or they could be assisted or regularized by
interactions with a companion.

The wind geometry, shaped by rotation and/or a companion star, may
also affect long-term spectroscopic variations.  Feast et al.\ (2001)
found in old spectrograms that $\eta$ Car was continuously in a
low-excitation (``event'') state for many years before 1920.  Most
likely the wind was then quasi-spherical as in Figure 18$b$.  The
large amount of mass ejected in $\eta$ Car's 19th century eruptions
(see Humphreys et al.\ 1999) represented most of the star's prior
radius, so we {\it expect\/} the surface of $\eta$ Car to have been
rotating very slowly a century ago.  Without rotation, there was no
reason for the wind to have the modern non-spherical shape shown in
Figure 18$a$; in particular there would not have been a lower-density
zone near the equatorial plane which today allows UV photons to
escape.  Therefore, no matter whether the hard UV comes from the
primary or from a secondary star, such radiation could have been
absorbed within the dense primary wind at low latitudes -- hence no
high-excitation emission lines then.  Later, we must suppose, the
surface rotation rate gradually increased for the reason noted above,
until it became sufficient to establish the current geometry and
high-excitation spectrum sometime around 1940 when a sudden
brightening occurred (de Vaucouleurs \& Eggen 1952; Thackeray 1953;
Gaviola 1953).  This coincidence suggests that the presence of
spectroscopic events is linked to changes in the primary star, which
dominates the observed optical flux.

Although $\eta$ Car is an extreme case, rotation-induced wind
structure and angular momentum diffusion may play important roles
for mass ejections from other Luminous Blue Variables.  This is
not a new revelation, but our latitude-dependent study of $\eta$
Car suggests the possibilities quite vividly.

\scriptsize
\acknowledgements

The authors benefitted from interesting and informative conversations
with Stan Owocki, Joe Cassinelli, and Roberta Humphreys.  An anonymous
referee gave several suggestions that improved the presentation of the
paper.  N.S. is grateful for the support of a NASA/GSRP fellowship
from Goddard Space Flight Center.  Additional support was provided by
NASA through grant number GO-8327 (P.I.: Davidson) from the Space
Telescope Science Institute, which is operated by the Association for
Research in Astronomy, Inc., under NASA contract NAS 5-26555, and
through STIS GTO resources and funding (P.I.: Gull).


\begin{table}
\caption{Measured parameters for H$\alpha$ in STIS spectra (March 2000)\tablenotemark{a}}
\scriptsize
\begin{tabular}{lcccccl}
\tableline\tableline 

Position &Width	   &$\Delta v$ &Latitude &$v_{\infty}$ &$v_{pcyg}$ &Comments \\

(arcsec) &(arcsec) &km s$^{-1}$ &(deg)	 &km s$^{-1}$   &km s$^{-1}$  &        \\

\tableline 

+7.3  &0.44 &785 &53 &--560    &--435    &NW lobe \\
+5.4  &0.57 &500 &45 &--475    &--410\tablenotemark{c}    &NW lobe \\
+3.3  &1.00 &350 &35 &\nodata  &--210\tablenotemark{b}  &fan, no P Cyg abs. \\
+2.6  &0.38 &220 &28 &\nodata  &--150\tablenotemark{b}  &fan, no P Cyg abs. \\
star  &0.15 &0   &49 &--540    &--410\tablenotemark{c}    &central star \\
--1.0 &0.51 &70  &56 &--640    &--430\tablenotemark{c}    &SE lobe \\
--1.5 &0.33 &105 &61 &--630    &--450\tablenotemark{c}    &SE lobe \\
--2.5 &0.63 &60  &66 &--700    &--460\tablenotemark{c}    &SE lobe, shadow \\
--3.0 &0.38 &100 &69 &--760    &--480    &SE lobe \\
--3.8 &0.38 &125 &73 &--780    &--480    &SE lobe \\
--4.3 &0.38 &125 &77 &--875    &--480    &SE lobe \\
--5.2 &0.38 &165 &83 &--1000   &--550    &SE lobe \\
--5.8 &0.25 &185 &86 &--1150   &--590    &SE lobe, polar axis \\
--7.0 &0.51 &270 &82 &--1050   &--540    &SE lobe \\
--8.0 &0.38 &350 &74 &--870    &--460    &SE lobe \\
NE 5\tablenotemark{d}  &1.00 &195 &0  &--750    &--440    &equatorial ejecta \\

\tableline
\end{tabular}
\tablenotetext{a}{See text in \S 2 for a description of the various
columns.}
\tablenotetext{b}{At these offset positions, $v_{pcyg}$ is the
velocity of the narrow Balmer absorption seen in the NW lobe.}
\tablenotetext{c}{At these offset positions, $v_{pcyg}$ was measured
for the H$\beta$ line, since H$\alpha$ velocities are ambiguous due to
contamination from [N~{\sc ii}] emission.  For H$\alpha$ and H$\beta$,
$v_{\infty}$ and $v_{pcyg}$ were the same to within $\lesssim$10 km
s$^{-1}$ in March 2000.}
\tablenotetext{d}{The spectrum of the NE5 offset position does not
fully sample the 1$\arcsec \times 1\arcsec$ box in Figure 1.  Rather,
adjacent slit positions were separated by 0$\farcs$25 intervals with a
52$\arcsec$$\times$0$\farcs$1 aperture oriented at P.A.=332$\fdg$1.}
\end{table}


\begin{table}
\caption{Measured parameters for H$\alpha$ in STIS spectra (March 1998; February 1999)\tablenotemark{a}}
\scriptsize
\begin{tabular}{lcccccccl}
\tableline\tableline 

Position &Width &$\Delta v$ &Latitude &$v_{\infty}$(1998)
&$v_{pcyg}$(1998) &$v_{\infty}$(1999) &$v_{pcyg}$(1999) &Comments \\

(arcsec) &(arcsec) &km s$^{-1}$ &(deg) &km s$^{-1}$ &km s$^{-1}$ &km
s$^{-1}$ &km s$^{-1}$ & \\

\tableline 

+7.4  &0.65 &930 &56 &--660 &--520 &\nodata &\nodata &NW lobe \\
+6.0  &1.84 &710 &49 &--650 &--500 &--480   &--360   &NW lobe \\
+4.6  &0.62 &450 &43 &\nodata  &\nodata &\nodata &--300\tablenotemark{b}    &NW lobe \\
+3.2  &0.95 &225 &36 &\nodata &--170\tablenotemark{b} &\nodata &--170\tablenotemark{b} &fan, no P Cyg abs. \\
+2.0  &0.85 &60  &31 &--750 &--580 &\nodata &\nodata &fan, no P Cyg abs. \\
star  &0.15 &0   &49 &--680 &--520 &--510\tablenotemark{c}    &--460\tablenotemark{c}  &central star \\
--1.1 &0.29 &70  &55 &--740 &--590 &\nodata &\nodata &SE lobe \\
--1.6 &0.47 &75  &60 &--850\tablenotemark{c} &--540  &--580\tablenotemark{c}  &--460  &SE lobe \\
--2.3 &0.55 &75  &64 &--780\tablenotemark{c} &--550  &\nodata  &--500\tablenotemark{c}  &SE lobe \\
--2.9 &0.70 &90  &67 &--780\tablenotemark{c} &--540  &\nodata  &--500\tablenotemark{c}  &SE lobe \\
--3.8 &0.57 &100 &71 &--610 &--480 &--580 &--460 &SE lobe \\
--4.4 &0.37 &145 &73 &--640 &--490 &--640 &--490 &SE lobe \\
--5.0 &0.49 &165 &76 &--650 &--500 &--680 &--500 &SE lobe \\
--5.7 &0.52 &192 &79 &--640 &--500 &--690 &--500 &SE lobe \\
--6.9 &0.52 &270 &78 &--620 &--480 &--650 &--480 &SE lobe \\
--7.5 &0.37 &345 &75 &--590 &--480 &--580 &--450 &SE lobe \\
--7.9 &0.39 &400 &72 &--600 &--480 &--580 &--460 &SE lobe \\  \tableline

\end{tabular}

\tablenotetext{a}{Similar to Table 1, but offset positions and
latitudes differ here because of the slit orientation used (see Figure
1).}  \tablenotetext{b}{At these offset positions, $v_{pcyg}$ is the
velocity of the narrow Balmer absorption seen in the NW lobe.}
\tablenotetext{c}{Poorly defined or contaminated by emission.}
\end{table}


\newpage

\begin{figure}
\epsscale{0.4}
\plotone{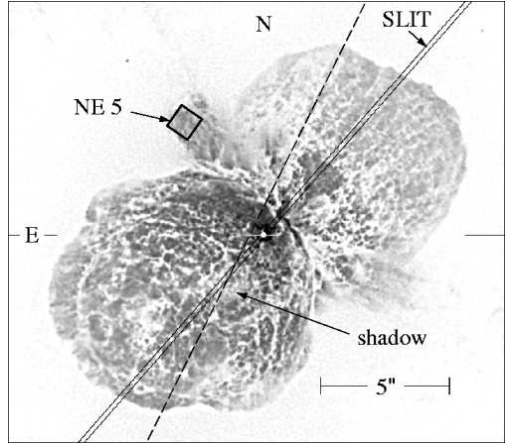}
\caption{\scriptsize Orientation and size of the STIS slit used for the March 2000
observations.  A second 1$\arcsec$ square aperture located 5$\arcsec$
NE of the star was used to extract the spectrum of one position in the
equatorial ejecta.  It is labeled ``NE 5'' and is discussed in \S 5.4.
The dashed line shows the orientation of the 0$\farcs$1-wide slit used
in March 1998 and February 1999 (see \S 6), which was intentionally
offset from the star by 0$\farcs$4.  The dark spot labeled ``shadow''
is discussed in \S 4.  The {\it HST}/WFPC2 image shown here was
obtained in 1995 (see Morse et al.\ 1998), but its scale has been
adjusted slightly to allow for expansion since then.}
\end{figure}

\begin{figure}
\epsscale{1}
\plotone{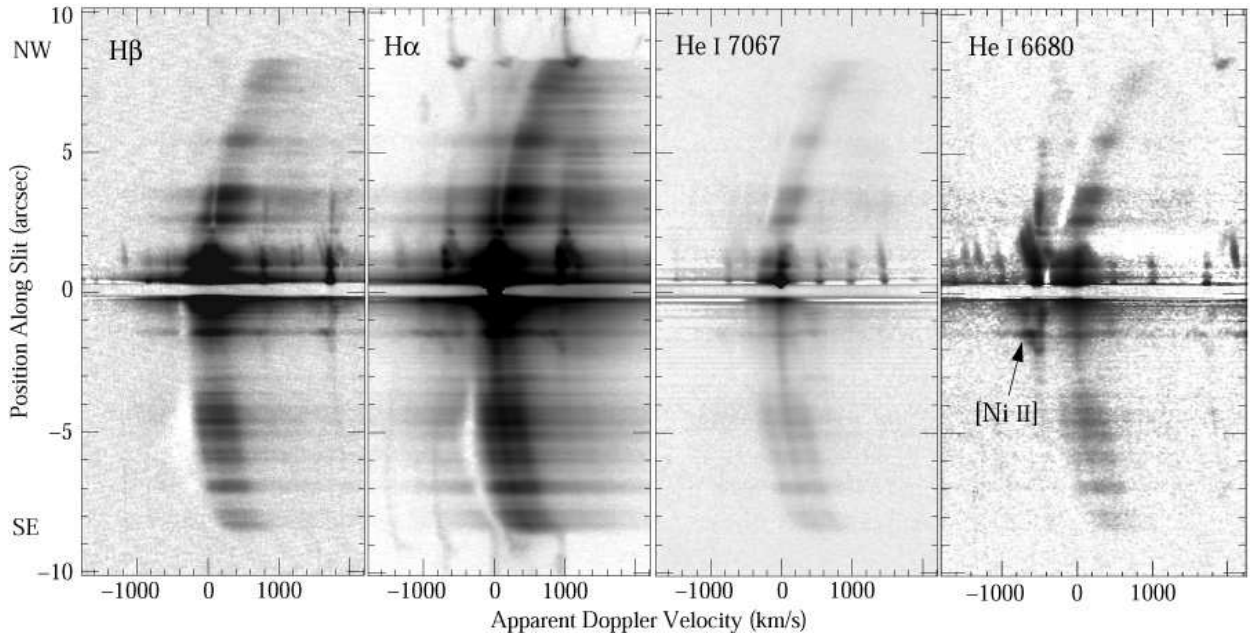}
\caption{\scriptsize Continuum-subtracted long-slit STIS spectra (March 2000) of a
few of the brightest reflected emission lines in $\eta$ Car's stellar
wind, using the 0$\farcs$2-wide long-slit aperture shown in Figure 1.}
\end{figure}

\begin{figure}
\epsscale{0.35}
\plotone{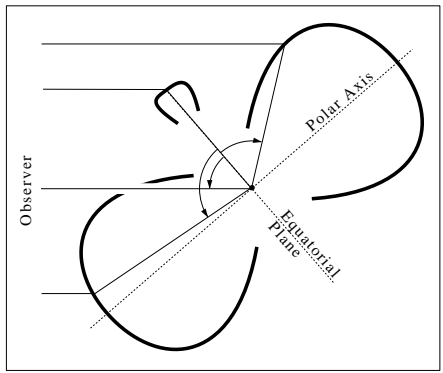}
\caption{\scriptsize Schematic diagram showing the relationship between position
and latitude for reflected light in the Homunculus Nebula. The
vertical axis corresponds to distance along the slit and the
horizontal axis is distance perpendicular to the plane of the sky.
The thick curved lines represent the ``compromise'' shape for the
reflecting surfaces of the bipolar lobes given by Davidson et al.\
(2001).  The arrows designating angular measure refer to latitudes
inferred for each line of sight.}
\end{figure}

\begin{figure}
\epsscale{0.6}
\plotone{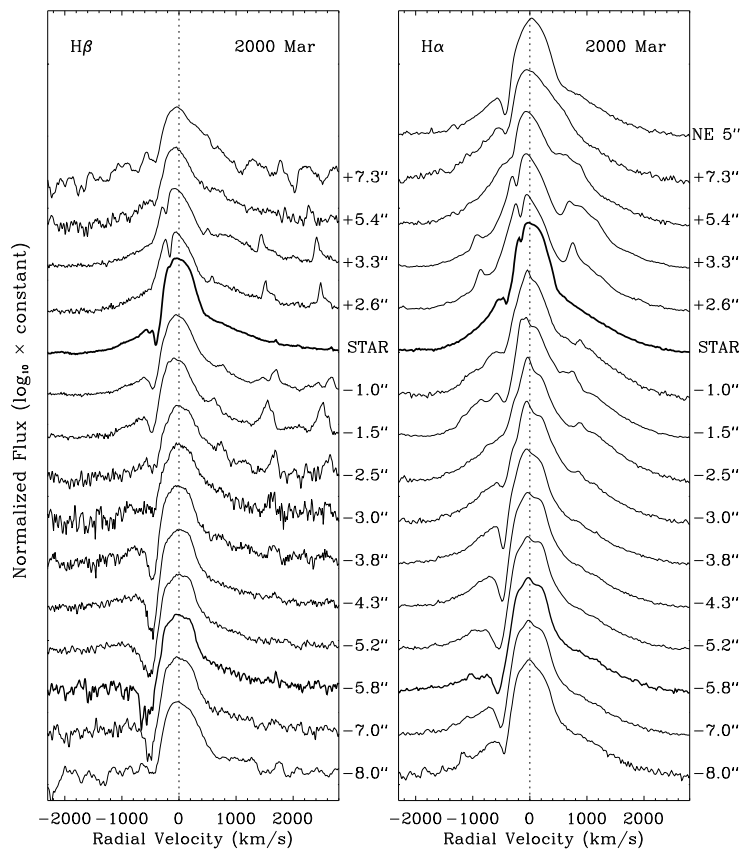}
\caption{\scriptsize Tracings of several different positions along the slit for
H$\beta$ and H$\alpha$, compared to the spectrum of the central star
(shown with a thick solid line).  Positional offsets for each line
profile are indicated at the right side of each panel.  The intensity
scale is logarithmic.  The narrow ``notch'' at $-$150 to $-$200 km
s$^{-1}$ seen at a few positions in the NW lobe is probably not due to
P Cyg absorption in the stellar wind, and will not be discussed in
this paper.  The tracing at $-$5$\farcs$8 is the closest to the polar
axis, and is drawn with a heavy line as well.}
\end{figure} 

\begin{figure}
\epsscale{0.8}
\plotone{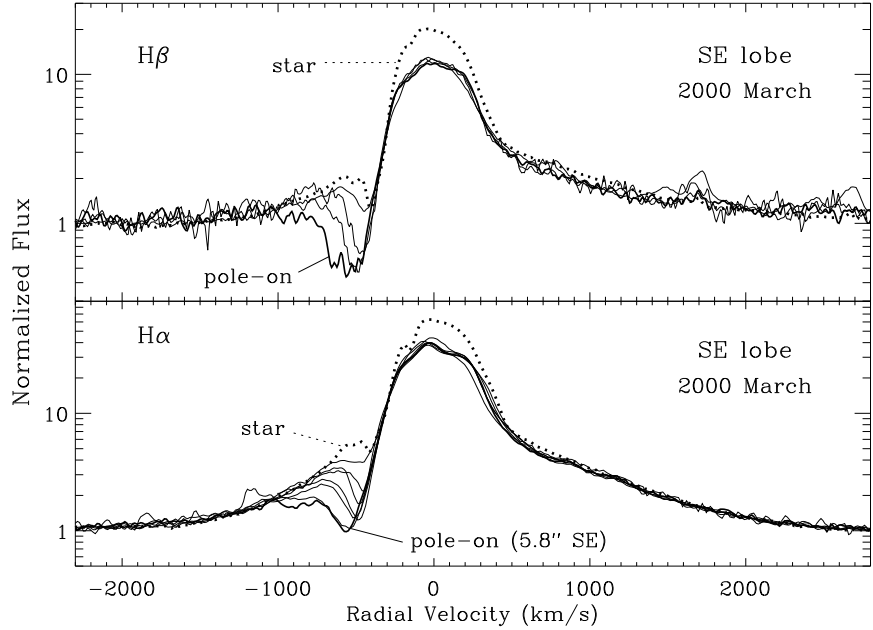}
\caption{\scriptsize Tracings of H$\beta$ and H$\alpha$ line profiles from Figure
4 for selected positions in the SE lobe and central star, normalized
to the same continuum level and superimposed on one another (log
intensity scale). Again, the narrow ``notch'' at $-$140 km s$^{-1}$ seen
in the central star's spectrum is probably not due to P Cyg absorption
in the stellar wind, and will not be discussed in this paper.  The
star's profile is taken from data obtained on 2000 March 20, as
described in the text.  These reflected profiles and those in other
figures have been shifted horizontally by the corresponding value of
$\Delta v$ in Table 1 for each tracing.}
\end{figure}

\begin{figure}
\epsscale{0.55}
\plotone{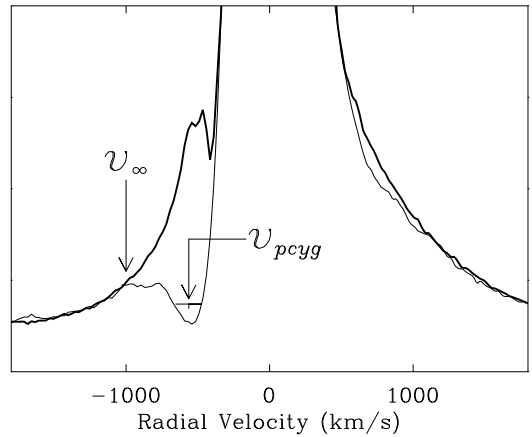}
\caption{\scriptsize H$\alpha$ profiles of the star (heavy line) and one position
in the SE lobe plotted on a linear intensity scale.  The arrows
indicate how we measured $v_{\infty}$ and $v_{pcyg}$. $v_{pcyg}$ is
the mean between the slopes of the absorption trough at half-depth.}
\end{figure}

\begin{figure}
\epsscale{0.55}
\plotone{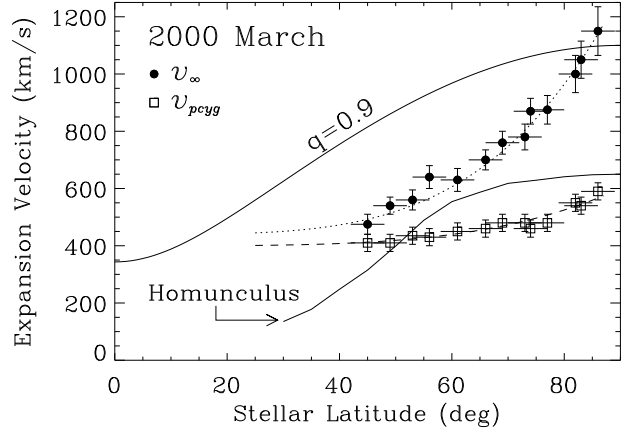}
\caption{\scriptsize Variation of $v_{\infty}$ and $v_{pcyg}$ for the H$\alpha$
line as a function of stellar latitude in $\eta$ Car's stellar wind.
The solid line labeled ``Homunculus'' corresponds to the expansion
velocities for the `compromise' shape of the polar lobes presented by
Davidson et al.\ (2001), if they were ejected in 1843.  The other
solid line is a curve for equation (1), with an arbitrary value of
$q = 0.9$.  The dotted and dashed lines show simple trends chosen
for comparison with the plotted values of $v_{\infty}$ and
$v_{pcyg}$.}
\end{figure}

\begin{figure}
\epsscale{0.4}
\plotone{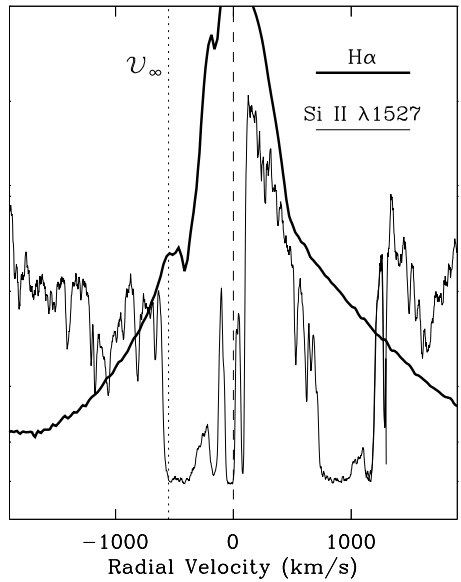}
\caption{\scriptsize H$\alpha$ line profile of the central star (thicker line)
observed on 20 March 2000, compared to the Si~{\sc ii} $\lambda$1527
line profile of the central star observed on 23 March 2000 with the
STIS FUV/MAMA detector.  Our method of measuring $v_{\infty}$ in
Balmer absorption profiles (dotted line) underestimates the true
$v_{\infty}$ (as measured in UV resonance lines) at a given latitude
in the wind by less than $\sim$50 km s$^{-1}$.}
\end{figure}

\begin{figure}
\epsscale{0.8}
\plotone{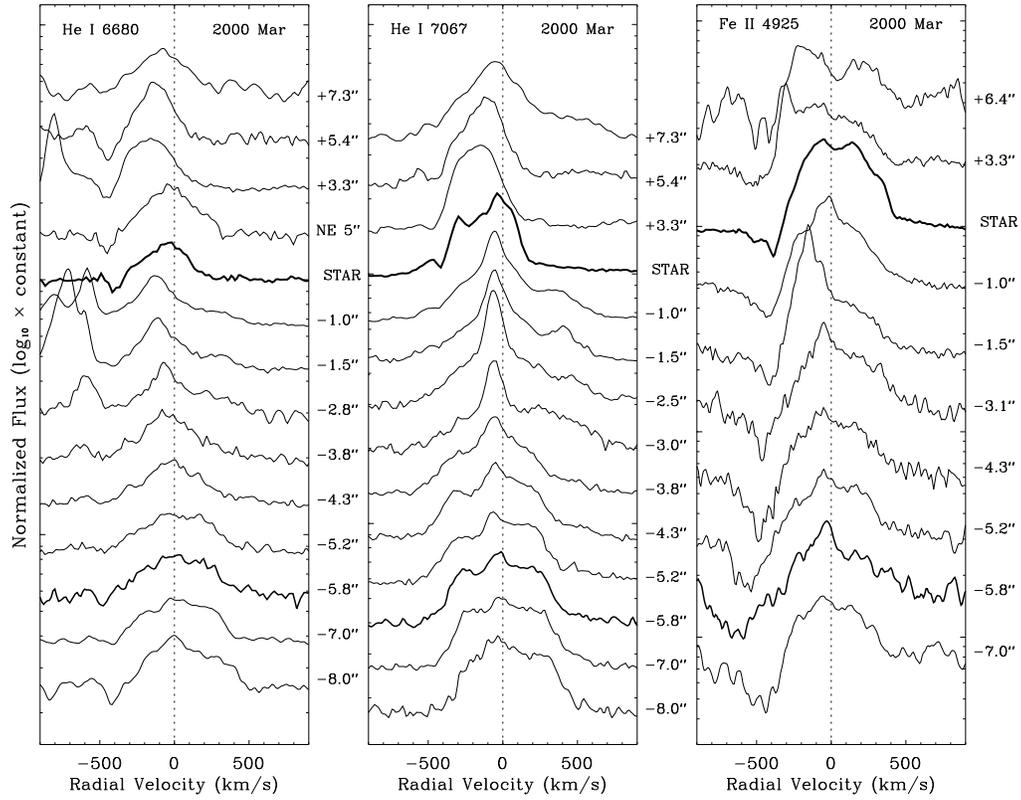}
\caption{\scriptsize Tracings of several different positions along the slit for
He~{\sc i} $\lambda$6680 and $\lambda$7067, and Fe~{\sc ii}
$\lambda$4925, compared to the spectrum of the central star (shown by
the thick solid line).  Positional offsets for each line profile are
indicated at the right side of each panel (log intensity scale).
Again, the tracing at $-$5$\farcs$8 is the closest to the polar axis,
and is drawn with a heavy line as well.}
\end{figure}

\begin{figure}
\epsscale{0.4}
\plotone{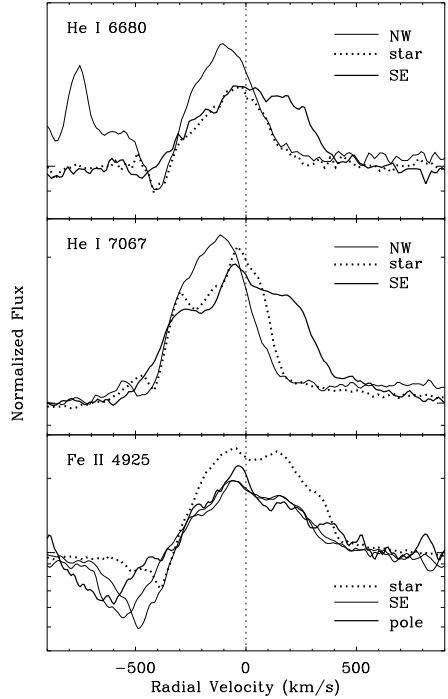}
\caption{\scriptsize Selected tracings in March 2000 (some are averages of adjacent tracings)
from Figure 6, normalized to the same continuum level and superimposed
on one another (logarithmic intensity scale).  ``NW'' and ``SE''
indicate the average reflected spectra from many positions in the NW
or SE polar lobe.  Tracings labeled ``pole'' are at $-$5$\farcs$8 (see
Table 1).  The NW sample for Fe~{\sc ii} $\lambda$4925 is excluded
because of contamination by narrow intrinsic emission.  Extra He~{\sc
i} $\lambda$6680 emission at velocities between $-$900 and $-$500 is due
to [Ni ~{\sc ii}] emission (see Figure 2).}
\end{figure}

\begin{figure}
\epsscale{0.5}
\plotone{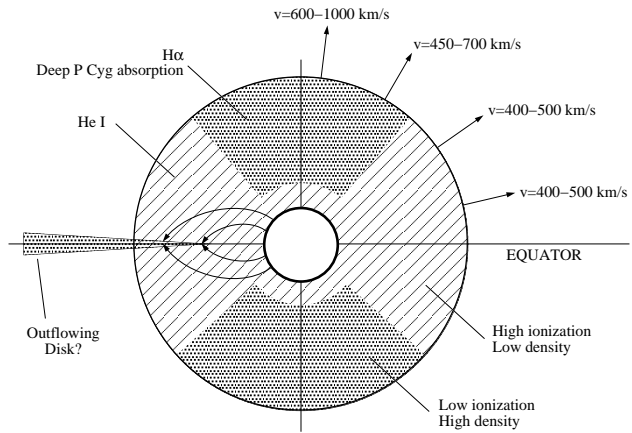}
\caption{\scriptsize Cartoon model for latitudinal structure in $\eta$ Car's
stellar wind during the normal high-excitation state in March 2000.}
\end{figure}

\begin{figure}
\epsscale{0.8}
\plotone{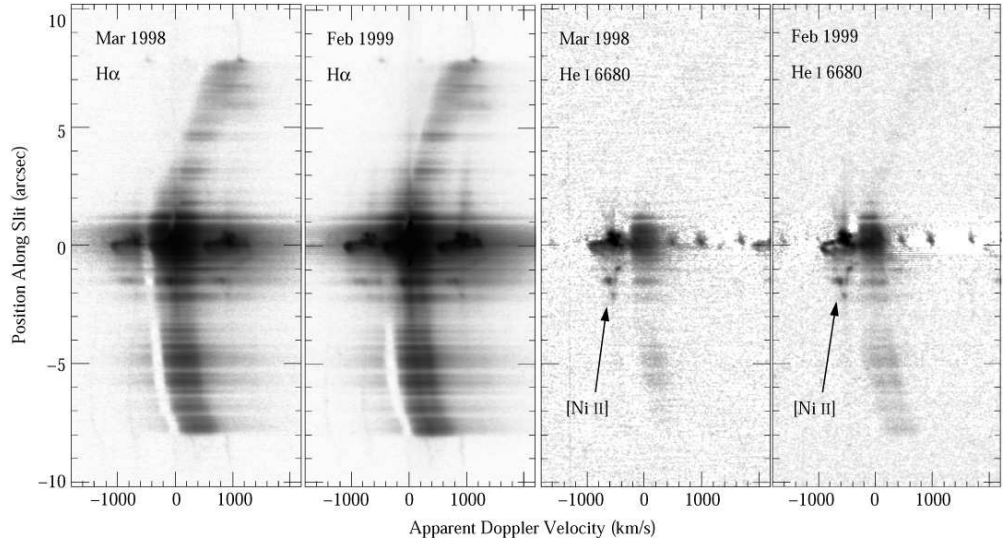}
\caption{\scriptsize Continuum-subtracted long-slit spectra of H$\alpha$ and
He~{\sc i} $\lambda$6680 in March 1998 and February 1999.  Compare
with the same emission lines in March 2000 shown in Figure 2.  Note,
however, a slight difference in position angle and placement of the
slit; these observations correspond to the dashed line in Figure 1.}
\end{figure}

\begin{figure}
\epsscale{0.8}
\plotone{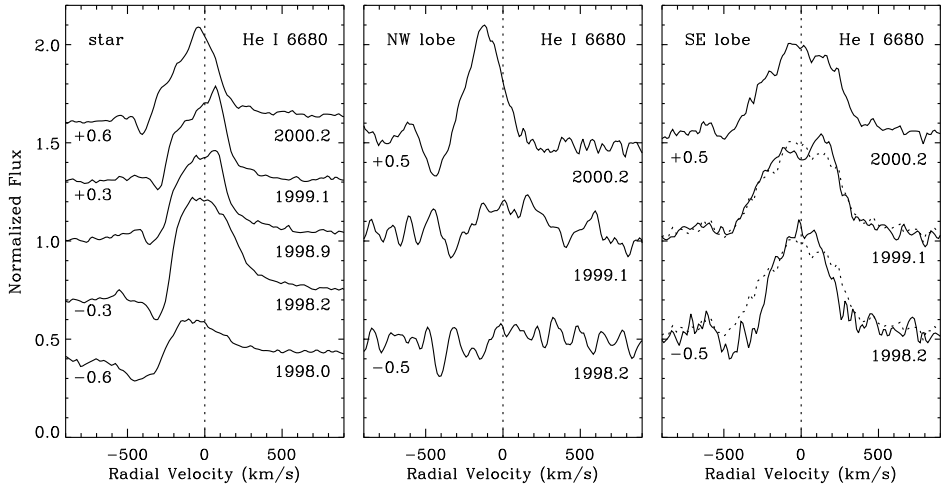}
\caption{\scriptsize Variability of the He~{\sc i} $\lambda$6680 line profile as
seen directly at the position of the central star (left), in the NW
polar lobe (center), and in the SE polar lobe (right).  The tracing in
the SE lobe for March 2000 is also superimposed on the SE lobe spectra
for earlier epochs with a dotted line.  A linear display scale is
used, and constant offsets are applied to the normalized spectra as
noted at the left in each panel.}
\end{figure}

\begin{figure}
\epsscale{0.5}
\plotone{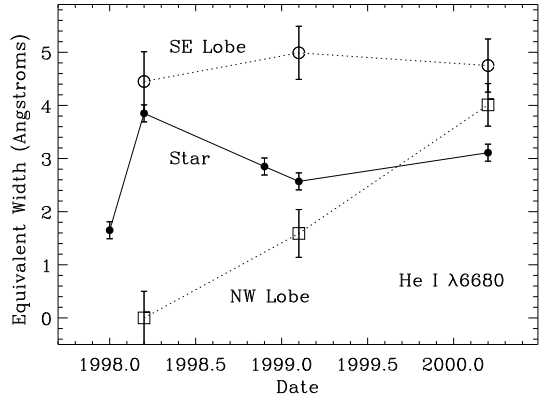}
\caption{\scriptsize Equivalent width of the He~{\sc i} $\lambda$6680 emission
line measured at several different epochs for the star, and reflected
spectra from the NW and SE polar lobes (see Figure 13).}
\end{figure}

\begin{figure}
\epsscale{0.65}
\plotone{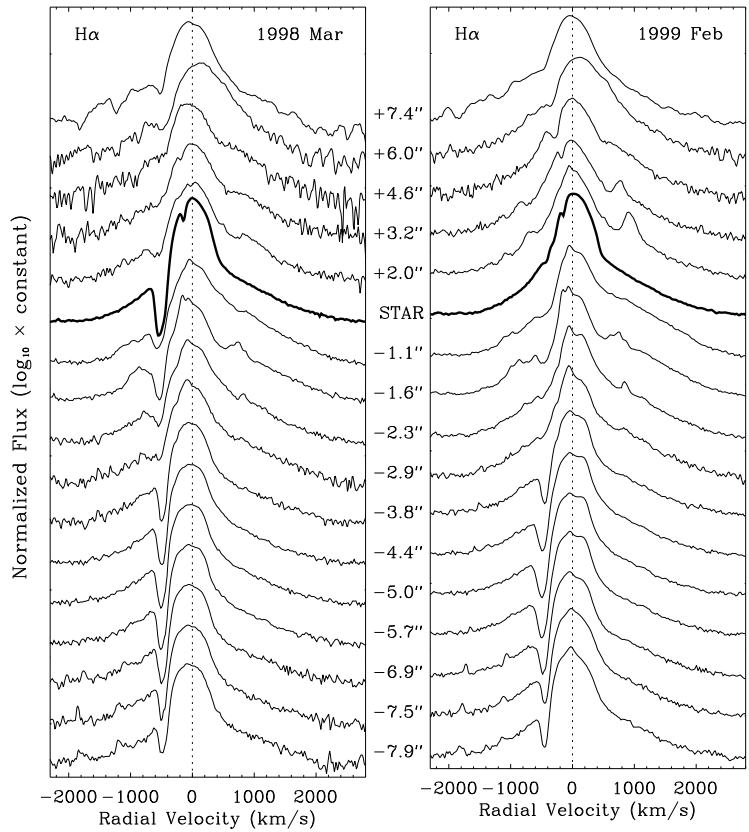}
\caption{\scriptsize Same as Figure 4 but for H$\alpha$ at two different times,
March 1998 (left) and February 1999 (right).  These tracings
correspond to the grayscale H$\alpha$ long-slit spectra in Figure 12,
with the slit P.A.\ oriented at 332$\arcdeg$ and offset 0$\farcs$4 NE
of the star.  Offset positions are the same for both dates and all are
indicated between the two panels.}
\end{figure}

\clearpage

\begin{figure}
\epsscale{0.6}
\plotone{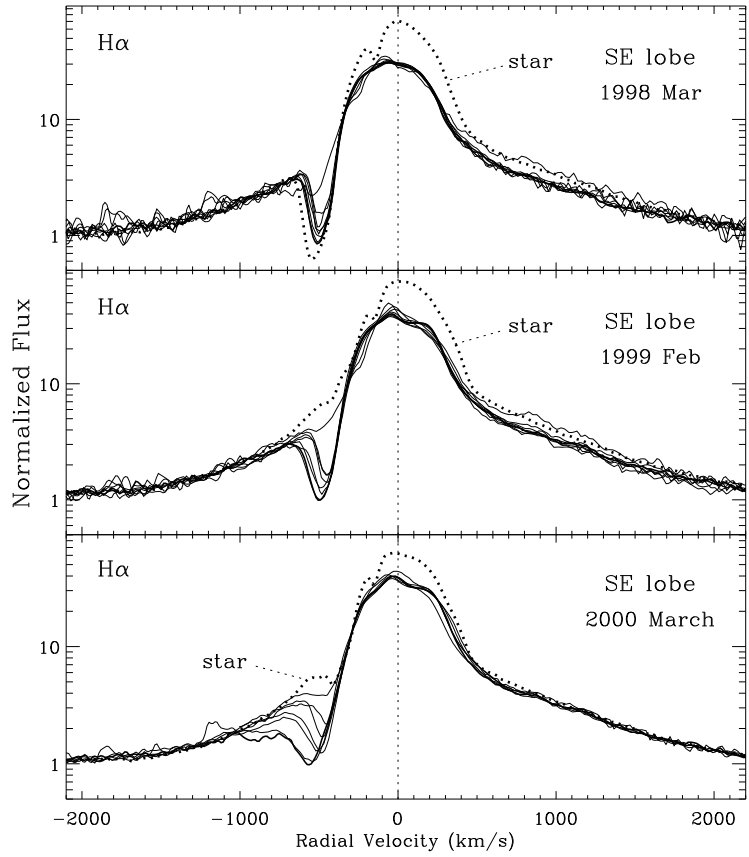}
\caption{\scriptsize Similar to Figure 5 but for H$\alpha$ at three different times,
March 1998, February 1999, and March 2000.  These tracings in 1998 and
1999 correspond to the grayscale H$\alpha$ long-slit spectra in Figure
13, with the slit P.A.\ oriented at 332$\arcdeg$ and offset 0$\farcs$4
NE of the star. March 2000 tracings are the same as in Figure 5. The
H$\alpha$ profile of the central star is shown by the dotted line for
each date.}
\end{figure}

\begin{figure}
\epsscale{0.4}
\plotone{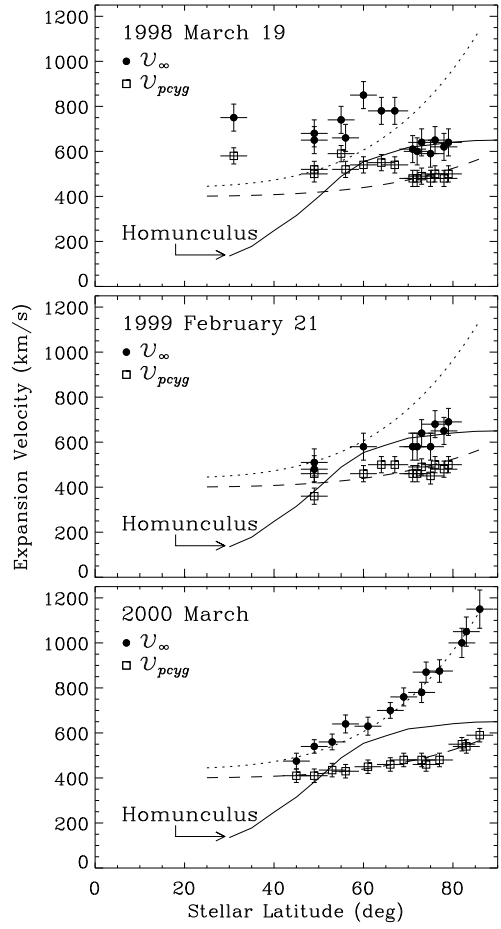}
\caption{\scriptsize Variation of $v_{\infty}$ and $v_{pcyg}$ for the H$\alpha$
line as a function of stellar latitude in $\eta$ Car's stellar wind.
Same as Figure 7, but for all three epochs, including March 1998 and
February 1999 (see Table 2), and March 2000 (Table 1).}
\end{figure}

\clearpage

\begin{figure}
\epsscale{0.5}
\plotone{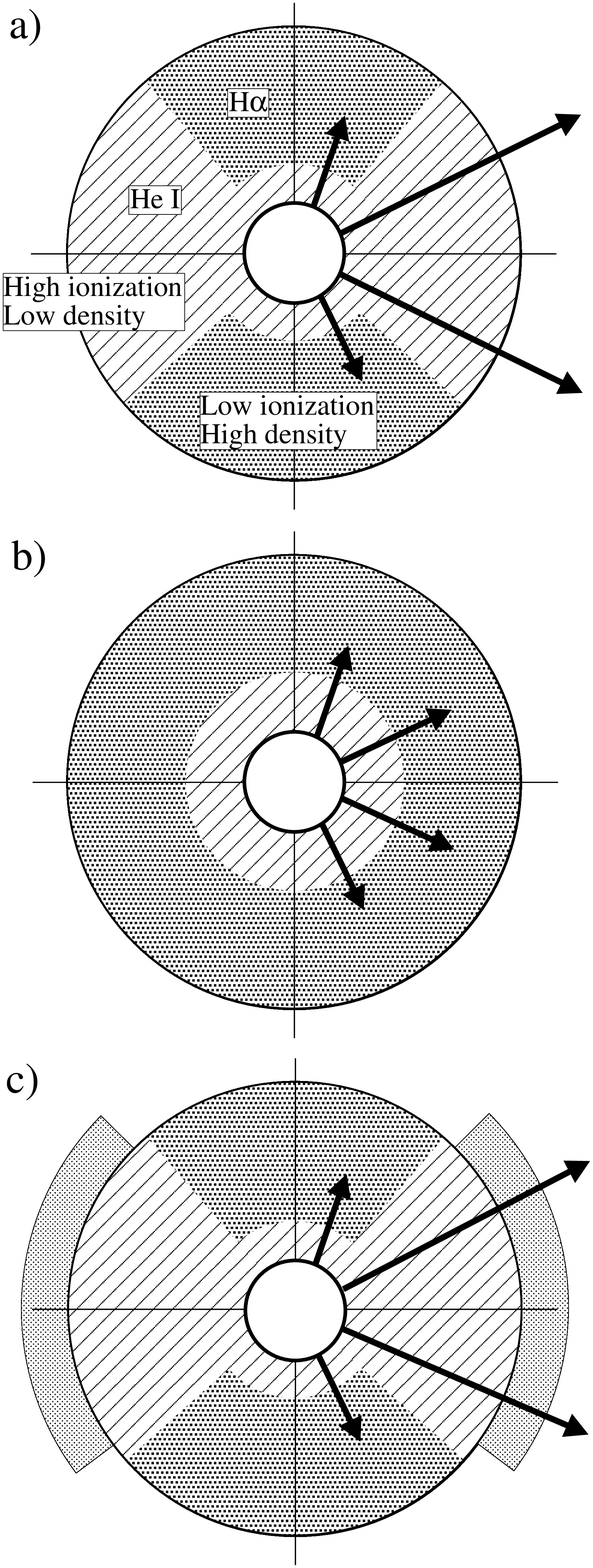}
\caption{\scriptsize Cartoon representing the structure of $\eta$ Car's wind
before (a), during (b), and after (c) a spectroscopic event as
indicated by changes in reflected emission line profiles in STIS
spectra.  Thick black arrows represent hard UV radiation.  See also
Figure 12.  The extra shaded regions to either side in panel (c)
represent a thinning shell ejected during the event.  This is not
intended to be a scale drawing in radius, and the real boundaries
between latitudinal zones are not so sharp.}
\end{figure}


\begin{references}

\reference{}Abraham, Z., \& Damineli, A.\ 1999, in ASP Conf. Ser. 179,
Eta Carinae at the Millenium, ed. J.A.\ Morse, R.M.\ Humphreys, \& A.\
Damineli (San Francisco: ASP), 263

\reference{}Balick, B., \& Matt, S.\ 2001, in ASP Conf. Ser. 242, Eta
Carinae and Other Mysterious Stars, ed. T.\ Gull, S.\ Johansson, \&
K.\ Davidson (San Francisco: ASP), 155

\reference{}Bidelman, W.P., Galen, T.A., \& Wallerstein, G.\ 1993,
PASP, 105, 785

\reference{}Bjorkman, J.E., \& Cassinelli, J.P.\ 1993, ApJ, 409, 429

\reference{}Castor, J.I., Abbott, D.C., \& Klein, R.I.\ 1975, ApJ, 195, 157

\reference{}Cranmer, S.R., \& Owocki, S.P.\ 1995, ApJ, 440, 308

\reference{}Corcoran, M.F., Rawley, G.L., Swank, J.H., \& Petre, R.\
1995, ApJ, 445, L121

\reference{}Corcoran, M.F., et al.\ 1998, ApJ, 494, 381

\reference{}Corcoran, M.F., Ishibashi, K., Swank, J.H., \& Petre, R.\
2001, ApJ, 547, 1034

\reference{}Currie, D.G., et al.\ 1996, AJ, 112, 1115

\reference{}Damineli, A.\ 1996, ApJ, 460, L49

\reference{}Damineli, A., Conti, P.S., \& Lopes, D.F.\ 1997, New
Astron., 2, 107

\reference{}Damineli, A., Stahl, O., Kaufer, A., Wolf, B., Quast, G.,
\& Lopes, D.F.\ 1998, A\&A SS, 133, 299

\reference{}Damineli, A., Kaufer, A., Wolf, B., Stahl, O., Lopes,
D.F., \& de Araujo, F.X.\ 2000, ApJ, 528, L101

\reference{}Davidson, K.\ 1999, in ASP Conf. Ser. 179, Eta Carinae at
the Millenium, ed. J.A.\ Morse, R.M.\ Humphreys, \& A.\ Damineli (San
Francisco: ASP), 304; see also p.\ 374

\reference{}Davidson, K., 2001, in ASP Conf. Ser. 242, Eta Carinae and
Other Mysterious Stars, ed. T.\ Gull, S.\ Johansson, \& K.\ Davidson
(San Francisco: ASP), 3

\reference{}Davidson, K., \& Humpherys, R.M.\ 1986, A\&A, 164, L7

\reference{}Davidson, K., \& Humphreys, R.M.  1997, ARAA, 35, 1

\reference{}Davidson, K., Dufour, R.J., Walborn, N.R., \& Gull, T.R.\
1986, ApJ, 305, 867

\reference{}Davidson, K., Ebbets, D., Weigelt, G., Humphreys, R.M.,
Hajian, A., Walborn, N.R., \& Rosa, M.  1995, AJ, 109, 1784

\reference{}Davidson, K., Ebbets, D., Johansson, S., Morse, J.A.,
Hamann, F.W., Balick, B., Humphreys, R.M., Weigelt, G., \& Frank, A.\
1997, AJ, 113, 335

\reference{}Davidson, K., Ishibashi, K., Gull, T.R., \& Humphreys,
R.M. 1999, in ASP Conf. Ser. 179, Eta Carinae at the Millenium,
ed. J.A.\ Morse, R.M.\ Humphreys, \& A.\ Damineli (San Francisco:
ASP), 227

\reference{}Davidson, K., Ishibashi, K., Gull, T.R., Humphreys,
R.M., \& Smith, N.\ 2000, ApJ, 530, L107

\reference{}Davidson, K., Smith, N., Gull, T.R., Ishibashi, K., \&
Hillier, D.J.\ 2001, AJ, 121, 1569

\reference{}de Vaucouleurs, G., \& Eggen, O.C.\ 1952, PASP, 64, 185

\reference{}Duncan, R.A., White, S.M., Nelson, G.J., Drake, S.A., \&
Kundu, M.R.\ 1995, ApJ, 441, L73

\reference{}Dwarkadas, V.V., \& Balick, B.\ 1998, AJ, 116, 829

\reference{}Dwarkadas, V.V., \& Owocki, S.P.\ 2002, preprint

\reference{}Ebbets, D.C., Walborn, N.R., \& Parker, J.W.\ 1997, ApJ,
489, L161

\reference{}Eddington, A.S. 1926, The Internal Constitution of the
Stars, sections 198--200

\reference{}Feast, M., Whitelock, P., \& Marang, F.\ 2001, MNRAS, 322,
741

\reference{}Frank, A., Balick, B., \& Davidson, K.\ 1995, ApJ, 441, L77

\reference{}Frank, A., Ryu, D., \& Davidson, K.\ 1998, ApJ, 500, 291

\reference{}Friend, D.B., \& Abbott, D.C.\ 1986, ApJ, 311, 701

\reference{}Gaviola, E.\ 1953, ApJ, 118, 234

\reference{} Glatzel, W.\ 1998, A\&A, 339, L5

\reference{}Gull, T.R., Ishibashi, K., Davidson, K., and the Cycle 7
STIS GO Team, 1999, in ASP Conf. Ser. 179, Eta Carinae at the
Millenium, ed. J.A.\ Morse, R.M.\ Humphreys, \& A.\ Damineli (San
Francisco: ASP), 144

\reference{}Gull, T.R., \& Ishibashi, K.\ 2001, in ASP Conf. Ser. 242,
Eta Carinae and Other Mysterious Stars, ed. T.\ Gull, S.\ Johansson,
\& K.\ Davidson (San Francisco: ASP), 59

\reference{}Guzik, J.A., Cox, A.N., \& Despain, K.M.\ 1999, in ASP
Conf. Ser. 179, Eta Carinae at the Millenium, ed. J.A.\ Morse, R.M.\
Humphreys, \& A.\ Damineli (San Francisco: ASP), 347

\reference{}Hamann, F., DePoy, D.L., Johansson, S., \& Elias, J.\
1994, ApJ, 422, 626

\reference{}Hamann, F., Davidson, K., Ishibashi, K., \& Gull,
T.R. 1999, in in ASP Conf. Ser. 179, Eta Carinae at the Millenium,
ed. J.A.\ Morse, R.M.\ Humphreys, \& A.\ Damineli (San Francisco:
ASP), 116

\reference{}Hillier, D.J., \& Allen, D.A.\ 1992, A\&A, 262, 153

\reference{}Hillier, D.J., Davidson, K., Ishibashi, K., \& Gull, T.R.\ 
2001, ApJ, 553, 837

\reference{}Humphreys, R.M., Davidson, K., \& Smith, N.\ 1999, PASP,
111, 1124

\reference{}Ignace, R., Cassinelli, J.P., \& Bjorkman, J.E.\ 1996,
ApJ, 459, 671

\reference{}Ishibashi, K., Corcoran, M.F., Davidson, K., Swank, J.H.,
Petre, R., Drake, S.A., Damineli, A., \& White, S.\ 1999, ApJ, 524,
983

\reference{}Ishibashi, K.\ 1999, in ASP Conf. Ser. 242, Eta Carinae
and Other Mystrious Stars, ed.\ T.\ Gull, S.\ Johansson, \& K.\ 
Davidson (San Francisco: ASP), 53

\reference{}Ishibashi, K., Gull, T.R., \& Davidson, K.\ 2001, in ASP
Conf. Ser. 242, Eta Carinae and Other Mysterious Stars, ed.\ T.\ Gull,
S.\ Johansson, \& K.\ Davidson (San Francisco: ASP), 71

\reference{}Johansson, S., \& Hamann, F. 1993, Physica Scripta, T47,
157

\reference{}Johansson, S., Leckrone, D.S., \& Davidson, K. 1998, in ASP
Conf.\ Ser.\ 143, The Scientific Impact of the GHRS, ed.\ J.C.\
Brandt, C.C.\ Petersen, \& T.B.\ Ake (San Francisco: ASP), 155

\reference{}Johansson, S., \& Letokhov, V.S.\ 2001, A\&A, 378, 266

\reference{}Kippenhahn, R., \& Weigert, A.  1990, Stellar Structure and
Evolution (Berlin: Springer-Verlag), section 42.2

\reference{}Langer, N., Garcia-Segura, G., \& Mac Low, M.M.\ 1999, ApJ,
520, L49

\reference{}Lamers, H.J.G.L.M., \& Pauldrach, A.W.A.\ 1991, A\&A, 244, L5

\reference{}Lamers, H.J.G.L.M., \& Cassinelli, J.P.\ 1999,
``Introduction to Stellar Winds'' (Cambridge: Cambridge University
Press)

\reference{} Lee, U., Saio, H., \& Osaki, Y.\ 1991, MNRAS, 250, 432

\reference{}L\'{o}pez, J.A., \& Meaburn, J.\ 1986, RevMexAA, 13, 27

\reference{}Maeder, A.\ 1999, A\&A, 347, 185

\reference{}Maeder, A., \& Desjacques, V.\ 2001, A\&A, 372, L9

\reference{}Maeder, A., \& Meynet, G.\ 2000, A\&A, 361, 159

\reference{}McGregor, P.J., Rathborne, J.M., \& Humphreys, R.M.\ 1999,
in ASP Conf. Ser. 179, Eta Carinae at the Millenium, ed. J.A.\ Morse,
R.M.\ Humphreys, \& A.\ Damineli (San Francisco: ASP), 236

\reference{}Meaburn, J., Wolstencroft, R.D., \& Walsh, J.R.\ 1987,
A\&A, 181, 333

\reference{}Meaburn, J., Gehring, G., Walsh, J.R., Palmer, J.W.,
Lopez, J.A., Bruce, M., \& Raga, A.C.\ 1993, A\&A, 276, L21

\reference{}Morse, J.A., Davidson, K., Bally, J., Ebbets, D., Balick,
B., \& Frank, A.  1998, AJ, 116, 2443

\reference{}Morse, J.A., Humphreys, R.M., \& Damineli, A.\ (eds.)
1999, ASP Conf. Ser. 179, Eta Carinae at the Millenium (San Francisco:
ASP)

\reference{}Morse, J.A., Kellogg, J.R., Bally, J., Davidson, K.,
Balick, B., \& Ebbets, D.\ 2001, ApJ, 548, L207

\reference{}Najarro, F., Hillier, J.D., \& Stahl, O.\ 1997, A\&A, 326, 1117

\reference{}Owocki, S.P.\ 2002, private communication

\reference{}Owocki, S.P., Cranmer, S.R., \& Blondin, J.M.\ 1994, ApJ,
424, 887

\reference{}Owocki, S.P., Cranmer, S.R., \& Gayley, K.G.\ 1996, ApJ,
472, L115

\reference{}Owocki, S.P., Gayley, K.G., \& Cranmer, S.R.\ 1998, in ASP
Conf. Ser. 131, Boulder Munich {\sc ii}: Properties of Hot Luminous
Stars, ed. I.D.\ Howarth (San Francisco: ASP), 237

\reference{}Owocki, S.P., \& Gayley, K.G.\ 1997, in ASP Conf. Ser. 120,
Luminous Blue Variables: Massive Stars in Transition, ed. A.\ Nota \&
H.J.G.L.M. Lamers (San Francisco: ASP), 121

\reference{}Pauldrach, A.W.A., \& Puls, J.\ 1990, A\&A, 237, 409

\reference{}Pittard, J.M., \& Corcoran, M.F.\ 2002, A\&A, 383, 636

\reference{}Puls, J.\ 1996, A\&A, 305, 171

\reference{}Shaviv, N.J.\ 2000, ApJ, 532, L137

\reference{}Smith, N., \& Gehrz, R.D.\ 1998, AJ, 116, 823

\reference{}Smith, N., Gehrz, R.D., \& Krautter, J.\ 1998, AJ, 116, 1332

\reference{}Smith, N., Gehrz, R.D., \& Krautter, J.\ 1999, in ASP
Conf. Ser. 179, Eta Carinae at the Millenium, ed. J.A.\ Morse, R.M.\
Humphreys, \& A.\ Damineli (San Francisco: ASP), 31

\reference{}Smith, N., \& Gehrz, R.D.\ 2000, ApJ, 529, L99

\reference{}Smith, N., Morse, J.A., Davidson, K., \& Humphreys, R.M.\
2000, AJ, 120, 920

\reference{}Smith, N., \& Davidson, K.\ 2001, ApJ, 551, L101

\reference{}Smith, N.\ 2001, in ASP Conf. Ser. 242, Eta Carinae and
Other Mysterious Stars, ed. T.\ Gull, S.\ Johansson, \& K.\ Davidson
(San Francisco: ASP), 81

\reference{}Smith, N.\ 2002, MNRAS, 337, 1252

\reference{}Smith, N., Gehrz, R.D., Hinz, P.M., Hoffmann, W.F.,
Mamajek, E.E., Meyer, M.R., \& Hora, J.L.\ 2002, ApJ, 567, L77

\reference{}Stothers, R.B.\ 2000, ApJ, 530, L103

\reference{}Thackeray, A.D.\ 1953, MNRAS, 113, 237

\reference{}Thackeray, A.D.\ 1961, Observatory, 81, 99

\reference{}ud-Doula, A., \& Owocki, S.P.\ 2002, preprint

\reference{}Viotti, R., Rossi, L., Cassatella, A., Altamore, A., \&
Baratta, G.B.\ 1989, \apjs, 71, 983

\reference{}Visvanathan, N.\ 1967, MNRAS, 135, 275

\reference{}von Zeipel, H.\ 1924, MNRAS, 84, 665

\reference{}Walborn, N.R., \& Liller, M.\ 1977, ApJ, 211, 181

\reference{}Washimi, H., \& Shibata, S.\ 1993, MNRAS, 262, 936

\reference{}White, S.\ 2002, private communication

\reference{}Whitelock, P.A., Feast, M.W., Koen, C., Roberts, G., \&
Carter, B.S.\ 1994, MNRAS, 270, 364

\reference{}Wolf, B., Kaufer, A., Stahl, O., \& Damineli, A.\ 1999, in
ASP Conf. Ser. 179, Eta Carinae at the Millenium, ed. J.A.\ Morse,
R.M.\ Humphreys, \& A.\ Damineli (San Francisco: ASP), 243

\reference{}Wolf, B., Stahl, O., \& Fullerton, A.W. (eds.) 1998,
Variable and Non-spherical Stellar Winds in Luminous Hot Stars
(Heidelberg: Springer)

\reference{}Zanella, R., Wolf, B., \& Stahl, O.\ 1984, A\&A, 137, 79

\reference{}Zethson, T., Johansson, S., Davidson, K., Humphreys, R.M.,
Ishibashi, K., \& Ebbets, D.\ 1999, A\&A, 344, 211

\reference{}Zickgraf, F.J., Wolf, B., Leitherer, C., Appenzeller, I.,
\& Stahl, O.\ 1986, A\&A, 163, 119


\end{references}
\end{document}